\documentclass[a4paper,10pt]{article}
\usepackage{amsmath, amssymb, amscd,bbm}
\usepackage[all]{xy}
\usepackage{slashed}

\begin{document}

\renewcommand{\thefootnote}{\fnsymbol{footnote}}

\begin{titlepage}
\begin{center}
\hfill LTH 894 \\
\vskip 10mm

{\Large
{\bf 
Euclidean Actions, Instantons, Solitons and Supersymmetry
}}

\vskip 10mm

\textbf{T. Mohaupt}\footnote{\textsf{Thomas.Mohaupt@liv.ac.uk}}
\textbf{and K. Waite}\footnote{\textsf{K.Waite@liverpool.ac.uk}}

\vskip 4mm

Theoretical Physics Division\\
Department of Mathematical Sciences\\
University of Liverpool\\
Liverpool L69 7ZL, UK \\

\end{center}

\vskip .2in 

\begin{center} {\bf ABSTRACT} \end{center}
\begin{quotation} \noindent
Theories with axionic scalars admit three different Euclidean 
formulations, obtained by Wick rotation, Wick rotation combined
with analytic continuation of the axionic scalars, and Wick rotation
combined with Hodge dualization. We investigate the relation between
these formulations for a class of theories which contains the 
sigma models of $N=2$ vector multiplets as a special case. It is shown
that semi-classical amplitudes can be expressed equivalently using
the two types of axionic actions, while the Hodge dualized version 
gives a different value for the instanton action unless the 
integration constants associated with the axion fields are 
chosen in a particular way. With this choice the instanton action 
is equal to the mass of the soliton or black hole obtained by 
dimensional lifting with respect to time. For supersymmetric models
we use the Euclidean supersymmetry algebra to derive a Euclidean
BPS condition, and we identify a geometrical criterion which 
distinguishes BPS from non-BPS extremal solutions.
\end{quotation}

\vfill

\end{titlepage}

\eject

\renewcommand{\thefootnote}{\arabic{footnote}}
\setcounter{footnote}{0}

\section{Introduction}

Instantons and solitons
provide important analytical tools to
study non-per\-tur\-ba\-tive effects of field and string theories. Often
they can be related to one another via dimensional reduction over time. One application
of this link is to generate stationary solutions, for 
example black holes, by lifting solutions of the time-reduced 
theory \cite{Breitenlohner:1987dg,Stelle:1998xg}. 
In the context of supergravity and string theory
instantons, or $(-1)$-branes, can act as seed solutions to generate
all the solitons, or $p$-branes with $p\geq 0$. Euclidean actions
constructed via reduction over time are in general not positive definite,
which raises the question whether their solutions, while definitely
useful for constructing solitons via lifting, can really be interpreted
as instantons. A related problem  is that upon substituting the
`candidate instanton' into the Euclidean action one obtains zero instead
of a finite positive value needed for a consistent semi-classical 
approximation. These problems are not tied to dimensional 
reduction over time, but always occur when 
looking for instanton solutions supported by 
scalars in more than one dimension. The reason is that
Derrick's theorem forbids purely scalar `instantons' (non-constant finite
action solution) and `solitons' (non-constant
time-independent finite energy solutions) in 
two or more space-like dimensions \cite{Derrick,Erice}.\footnote{Exceptions
arise for models 
without a scalar potential if the 
Laplacian has non-trivial zero modes. The world sheet 
instantons of string theory are an example of this.}

It is well known that there is a class of axionic instanton solutions 
which circumvents this no go theorem, in particular 
axionic wormhole-type solutions 
\cite{GiddingsStrominger:1988,Brown:1989df,Coleman:1989zu,BurgessKshirsagar:1989}, 
the 
D-instanton solution of type-IIB supergravity 
\cite{Gibbons:1995vg,Green:1997tv}
\footnote{The D-instanton
solution corresponds to flat Euclidean space in the Einstein frame, but
is a finite neck wormhole in the string frame.} 
and hypermultiplet and vector multiplet instanton solutions in $N=2$
string compactifications \cite{Behrndt:1997ch,Gutperle:2000sb,Gutperle:2000ve,Theis:2002er,Davidse:2003ww,deVroome:2006xu,EucIII}.
There are three different approaches to such
axionic instantons. The first makes use of the fact that 
like finite dimensional integrals, functional integrals
can be dominated by a complex rather than real saddle 
point \cite{Brown:1989df,Coleman:1989zu}. One
then has to show that there exists a complex 
saddle point which leads to a consistent semi-classical approximation 
to the functional integral one wishes to compute. The tunneling 
through a circular potential
barrier provides a simple quantum mechanical toy example 
for this type of solution \cite{Brown:1989df}.
Another approach is based on the
fact that Wick rotation and Hodge dualization do not commute and
uses a dual Euclidean action where the axions have been replaced by
antisymmetric tensor fields \cite{GiddingsStrominger:1988,Brown:1989df,BurgessKshirsagar:1989,Gibbons:1995vg,Green:1997tv}. This Euclidean action is 
positive 
definite and has real saddle points, which is not in contradiction
to Derrick's theorem because the tensor fields have a local 
gauge symmetry. At the level of individual amplitudes
the tensor formulation can be related
to the axionic formulation by a change of variables in
the functional integral. However, the Hilbert spaces of axionic and
tensor field theory are not isomorphic in general, \cite{BurgessKshirsagar:1989}
and the breaking of continuous axionic shift symmetries is not
straightforward to describe.
The third approach is to use an 
indefinite Euclidean action with inverted kinetic terms for 
the axions \cite{Gibbons:1995vg,Green:1997tv}. 
For such actions Derrick's theorem does not 
apply and real saddle points exist. Actions with inverted kinetic
terms for axions result
from dimensional reduction over time, but can also be constructed
by either applying Hodge dualization 
before and after the Wick rotation, or by using
a modified Wick rotation, where 
axionic fields are continued analytically to imaginary 
values 
\cite{Zumino:1977yh,Gibbons:1995vg, vanNieuwenhuizen:1996tv,Green:1997tv}.
One reason for using a modified Wick rotation for axions is 
Euclidean supersymmetry. 
Poincar\'e and Euclidean supersymmetry algebras
have different R-symmetry groups, which in turn implies different
geometries of their scalar target spaces. For four-dimensional 
$N=2$ vector multiplets it was already observed in \cite{Zumino:1977yh}
that the change 
\[
U(1)_R \times SU(2)_R \rightarrow SO(1,1)_R \times SU(2)_R
\]
in the abelian factor of the R-symmetry group
implies an analytical continuation of all vector multiplet
axions. 
It was shown in \cite{EucI,EucIII} that this can be 
understood geometrically as replacing the complex structure
of the scalar target space by a para-complex structure. 
Based on this observation, 
Euclidean variants of the special geometries of both
rigid and local vector multiplets have been formulated.

We remark in passing that it is perfectly consistent to use a Euclidean 
version of a supersymmetric theory which is not invariant
under the Euclidean supersymmetry, as long as one can derive
Euclidean Ward identities which give the supersymmetric Ward identities
upon continuation to Minkowski space \cite{Nicolai:1978}. 
Options for the Euclidean formulation of supersymmetric theories have
also been discussed 
\cite{vanNieuwenhuizen:1996tv,vanNieuwenhuizen:1996ip,Blau:1997pp,Belitsky:2000ii,Theis:2001ef}. 
In this paper we will not attempt to give a comprehensive 
discussion of the various existing approaches to Euclidean supersymmetry.
Instead we focus on the specific
properties of axionic instanton solutions for the Euclidean 
$N=2$ vector multiplets constructed in \cite{EucI,EucIII}, in 
particular the Euclidean BPS condition and its geometric interpretation.

The other main purpose of this paper is to clarify the status of indefinite
Euclidean actions for axions, whether
supersymmetric or not. 
We use the class of Euclidean
sigma models introduced in \cite{Mohaupt:2009iq}, which includes the scalar
sector of Euclidean $N=2$ vector multiplets as a special case. While it was
demonstrated in \cite{Mohaupt:2009iq} and \cite{Mohaupt:2010fk} that these models
have a rich set of solutions which can be lifted to extremal 
multi-centered and non-extremal single centered black hole solutions
in five dimensions, we are now focussing on the question whether
these solutions can be interpreted as instantons. As we will see 
this class of models is suitable for exploring the relations between
the three types of Euclidean actions introduced above, because
instanton solutions can be constructed in a systematic way.
Two constraints are imposed on the scalar metric:
(i) the metric must be determined by a potential,
a feature which generalizes the geometry of $N=2$ vector multiplets,
(ii) the continuous axionic shift symmetries must form an abelian 
group. This allows a simple lifting of the Euclidean theory 
with respect to time, and implies a simple and transparent
structure of the solutions, which helps us to investigate the
conceptual points we are interested in. 
These constraints admit
models with an arbitrary number of scalar fields which can be
treated in a universal
way, and do not require the scalar target space to be a 
symmetric or homogeneous space. This is a reasonable compromise
between generality and analytical control over solutions.

While the role of complex saddle points when using definite
axionic actions, the use of the Hodge dual tensor field
action as an alternative description of instantons, and the
use of indefinite axionic actions in order to generate solutions
by dimensional lifting have been discussed  
in detail in the literature, it has remained unclear which role
indefinite axionic actions play in the context of instantons
(as remarked in \cite{Chiodaroli:2009cz}).
For our class of models we find a simple answer
by re-investigating  the amplitude calculation of 
\cite{Coleman:1989zu,Chiodaroli:2009cz}: amplitudes can be expressed
both in terms of the definite and the indefinite action, and both
formulations are related by a change of variables which is nothing
but the analytical continuation of the axions which relates
the two types of actions. While in one
formulation the functional integral 
has an imaginary saddle point and one needs to integrate over
real fluctuations around it, in the other formulation the saddle 
point is real and the fluctuations are imaginary. Given that the
two actions are related by analytical continuation this might 
sound obvious, but, as we will see, the calculation of amplitudes
involves the careful consideration of boundary conditions and boundary
terms. In particular, since the naive bulk action vanishes when
evaluated on instantons, boundary terms are essential for obtaining
a consistent semi-classical approximation where instanton contributions
are exponentially suppressed. As we will see, the boundary
terms obtained are related by the same analytical continuation 
as the bulk actions, which is needed for both formulations being
equivalent. 
Moreover, the amplitude calculation is instructive because it shows
that for axionic instantons we cannot avoid to consider 
complex values of the axions. The situation is analogous to 
computing an integral over the real line which has a complex saddle 
point by contour deformation. Adopting a `complex point of view'
shifts the emphasis from the quest for `the' correct Euclidean 
action to the more appropriate one of choosing the appropriate 
`integration contour' for the semi-classical approximation of 
a given functional integral. A similar point of view is taken
in \cite{Eynard:2008yb} in the context of matrix models.

Since the dual tensor action is positive definite and
has real saddle points, one might think that one could abolish
the axionic actions altogether and only use the tensor
formulation. However, there are problems with this idea. In 
the supersymmetric context there are cases, notably the dilaton
multiplet in heterotic compactifications, where no appropriate
dual off shell supermultiplet exists \cite{Siebelink}. 
More generally, it is not clear
how to capture the breaking of continuous axionic shift symmetries
in the tensor formulation. The reason is that while axions have global
shift symmetries, tensor fields have local gauge symmetries, and therefore
it is not clear how the breaking of continuous shift symmetries, 
which is believed to be a genuine physical effect, could be captured
by a local effective scalar-tensor action. The program of finding the full
non-perturbative corrections to hypermultiplet moduli spaces (see
\cite{Saueressig:2007gi} for a review)  faces 
precisely this problem.
Therefore it is desirable to develop
methods for computing instanton effects which do not rely on the
tensor formulation.

This asymmetry between two theories which are completely equivalent
classically, or, 
more precisely, at the level of their equations of motion, indicates
that they are not fully equivalent as quantum field theories.  
This observation 
has already been made in a different context in 
\cite{BurgessKshirsagar:1989}, where the relation between the Hilbert
spaces of axionic and tensor field theories was investigated.
By taking space to be compact and applying Hodge decomposition to the
space of classical solutions, the precise relation between the Hilbert
spaces of the two theories was worked out. It turned out that the
Hilbert spaces are not isomorphic,
but differ in their zero mode parts. The zero modes of the axions,
while completely trivial in the classical theory, label superselection
sectors of the quantum theory. This has no analogue in the tensor theory,
which has a local gauge symmetry, and therefore the Hilbert space of the 
axionic theory is larger.\footnote{However one can embed the axionic 
Hilbert space into multiple copies of the tensor field Hilbert space. Thus any
problem in one theory can be reformulated within the other.} 
We encounter a different aspect of this non-equivalence 
(or `equivalence up to zero modes') when re-investigating the relation
between the semi-classical evaluations of the respective partition functions. 
Here we observe
that in the axionic formulation the instanton action explicitly
depends on the boundary values of the axions, which is a manifestation
of the breaking of continuous shift symmetries. In contrast,
the instanton action of the dual tensor field theory does not depend
explicitly on two-form potential, as required by local gauge symmetry.
As a consequence
the instanton actions of both pictures only agree if the boundary 
values of the axions, which are integration constants of the instanton 
solution, are adjusted in a particular way. With this adjustment the
instanton action is equal to the ADM mass of the black hole obtained
by lifting. 

In the last part of the paper we specialize to supersymmetric 
models. We show that the Euclidean BPS condition can be derived
from the Euclidean supersymmetry algebra and that it is related
by dimensional lifting to the BPS condition for a massive charged 
point-like BPS state. We also find a geometrical characterization
of the distinction between BPS and extremal non BPS solutions in 
terms of the para-complex geometry of the target space, which
generalizes to supersymmetric theories.

Throughout the paper we restrict ourselves to what we call extremal 
Euclidean solutions. The working definition
of `extremal' is a solution with vanishing energy momentum tensor.
Such solutions remain completely unmodified when coupling 
the theory to gravity. This observation can be used to construct
extremal black hole solutions in terms of a scalar sigma model
on flat Euclidean space. Moreover, in the 
dual tensor formulation such solutions saturate 
a Bogomol'nyi bound, and in the supersymmetric
case BPS solutions are extremal.

The paper is organised as follows. In Section 2 we introduce 
the class of four-dimensional sigma models that we will use
for our investigations. Besides the Minkowski signature version
two Euclidean versions are introduced, one with a definite, the
other with an indefinite action. We review the class of instanton 
solutions constructed in \cite{Mohaupt:2009iq}, which are complex saddle points
of the definite and real saddle points of the indefinite Euclidean
action. In Section 4 we adapt the calculation of semi-classical 
transition amplitudes of \cite{Coleman:1989zu,Chiodaroli:2009cz}
to our class of models and show that a consistent saddle point
approximation can be performed using either form of the Euclidean action,
and with identical result. In Section 5 we investigate the relation
to the third type of Euclidean action, where all axions have been 
dualized into tensor fields, and point out subtleties related to
the treatment of the axionic zero modes. In Section 6 we briefly
review how the instanton solutions can be lifted to five-dimensional
solitons or black holes, depending on whether the theory is coupled
to gravity or not, and compare the instanton action to the soliton and
black hole mass. In Section 7 we specialize to supersymmetric models.
The Euclidean BPS condition for purely scalar field configurations 
of $N=2$ vector multiplets is derived and shown to be related to 
the BPS condition of massive BPS states in one dimension higher.
We also find a relation between the BPS condition and the geometry
of the target space. We conclude in Section 8 where we identify
open questions and future lines of research.

\section{Instanton solutions}

\subsection{The Minkowski action}

We start with a theory of $2n$ scalars 
$\sigma^i$, $b^i$, $i=1, \ldots, n$ 
in four-dimensional Minkowski space $M$ with an action of the form
\begin{equation}
\label{MinkAct}
S_M[\sigma,b] = -\frac{1}{2} \int_M  d^4x  N_{ij}(\sigma) 
(\partial_\mu \sigma^i \partial^\mu \sigma^j + 
\partial_\mu b^i \partial^\mu b^j) \;,
\end{equation}
where $N_{ij}(\sigma)$ is positive definite, and $\mu=0,1,2,3$. 
This is a non-linear
sigma model with $n$ commuting isometries acting as shifts
$b^i \rightarrow b^i + C^i$. The corresponding Noether currents
and Noether charges are
\[
j_{\mu,i} = N_{ij}(\sigma) \partial_\mu b^j
\]
and 
\[
Q_i = \int_{x^0={\rm const}} d^3 {\bf x} j_{0,i}({\bf x},t)
\]
respectively.

The $2n$ real scalars can be combined into $n$ complex scalar
fields
\[
X^i = \sigma^i + i b^i \;,
\]
and the resulting line element on the scalar manifold ${\cal M}$
is Hermitean:
\[
ds^2_{\cal M} = N_{ij}(\sigma) (d\sigma^i d\sigma^j + d b^i d b^j) 
= N_{ij}(X+\bar{X}) dX^i d \bar{X}^j \;.
\]
Due to the shift symmetries, we can interpret ${\cal M}$ as the tangent
bundle $T{\cal S}$ of the manifold ${\cal S}$ 
parametrized by the scalars $\sigma^i$ alone,
with line element $ds^2_{\cal S} = N_{ij}(\sigma) d\sigma^i d\sigma^j$.

An important subclass, where solutions can be obtained
explicitly, are manifolds  ${\cal M}$ which are not only Hermitian,
but K\"ahler. This is equivalent to requiring that ${\cal S}$ is
Hessian \cite{AlekCor}. If $N_{ij}$ is a Hessian metric on ${\cal S}$
with Hesse potential ${\cal V}(\sigma)$,
\[
N_{ij} = \frac{\partial^2 {\cal V}}{\partial \sigma^i \partial \sigma^j} \;,
\]
then 
\[
K(X,\bar{X}) = K(X+\bar{X}) = 4 {\cal V}(\sigma(X+\bar{X}))
\]
is a K\"ahler potential for $ds^2_{\cal M} = N_{ij} d X^i d\bar{X}^j$:
\[
N_{ij} = \frac{\partial^2 K}{\partial X^i \partial \bar{X}^j} \;.
\]

\subsection{The definite  Euclidean action}

The standard  version 
of the Euclidean action is obtained by Wick rotation 
$x^0=-it$. Here $x^0$ denotes Minkowski time and $t$ denotes 
Euclidean time. The Wick rotation amounts to taking $t$ rather
than $x^0$ to be real.  The Euclidean action is 
\begin{equation}
\label{EucActDef}
S_E[\sigma,b] = \frac{1}{2} \int_E d^4 x  N_{ij}(\sigma) 
(\partial_m \sigma^i \partial^m \sigma^j 
+ \partial_m b^i \partial^m b^j) \;,
\end{equation}
where $m=1,2,3,4$ are Euclidean indices. Here $x^4=t$ is Euclidean 
time and if we want to emphasize this interpretation we use 
$t$ instead of 4 as an index.
The Euclidean action
is positive definite, $S_E \geq 0$, and takes its minimal 
value $S_E =0$ for constant scalars. The partition function
\[
Z = \int D\sigma D b e^{-S_E(\sigma, b)}
\]
is `damped', i.e. fluctuations around saddle points 
are suppressed, and we expect that the semi-classical 
approximation (saddle point approximation) is meaningful.
But by  Derrick's theorem \cite{Derrick,Erice}
there are no real saddle points.
The complex saddle points relevant for instanton effects
can be found by using a different type of Euclidean action.

\subsection{The indefinite scalar action}

Let us next discuss another choice for the Euclidean action,
where the Wick rotation is combined with an analytical continuation 
$b^i \rightarrow ib^i$ of the axionic scalars. To discuss the 
definite and indefinite version of the Euclidean action in parallel we will
use the notation $\beta^i = i b^i$ for the rotated axionic fields. 
Both Euclidean actions are related by analytic continuation, and 
it is useful to consider them as two different real forms of a complex
action, which arise by taking either $\beta^i$ or $b^i$ to be real.
The indefinite Euclidean action is
\begin{equation}
\label{EucActIndef}
\tilde{S}_E[\sigma,\beta] =
\frac{1}{2} \int_E d^4 x  N_{ij}(\sigma) 
(\partial_m \sigma^i \partial^m \sigma^j 
- \partial_m \beta^i \partial^m \beta^j) \;.
\end{equation}
The modified scalar manifold ${\cal E}$ obtained by rotating the
axions has the line element
\[
ds^2_{\cal E} = N_{ij}(\sigma) (d\sigma^i d\sigma^j - d \beta^i d \beta^j) \;.
\]
This has split signature $(n,n)$ so that the metric and the corresponding
action are indefinite. To understand the geometry underlying instanton 
solutions it is useful to note ${\cal E}$ carries a para-Hermitian 
metric, in analogy to the Hermitian metric on ${\cal M}$. In terms of 
local coordinates, we can combine the $2n$ real scalar fields $\sigma^i$ 
and $b^i$ into $n$ para-complex fields
\[
X^i = \sigma^i + e \beta^i
\]
where the para-complex unit satisfies
\[
e^2 = 1 \;,\;\;\;\bar{e} = - e \;.
\]
Para-complex geometry is in many respects analogous to complex 
geometry.\footnote{See \cite{EucI} for a comprehensive review.}
An almost para-complex structure $J$ is a tensor field of type $(1,1)$ which
satisfies
\[
J^2 = \mathbbm{1}
\]
and the additional requirement that $J$ has an equal number of eigenvalues
$+1$ and $-1$.\footnote{If the second requirement is dropped, then $J$ is an
almost product structure. } An almost para-complex structure is called 
a para-complex structure, if the analogue of the Nijenhuis tensor vanishes,
and this condition is equivalent to the existence of local para-complex
coordinates. If a line element takes the form
\[
ds^2_{\cal E} = N_{ij} dX^i d \bar{X}^j \;,
\]
then it is para-Hermitian.\footnote{To be able to define a fundamental two-form,
the para-complex structure is required to act as an anti-isometry. In 
the context of `doubled geometries', this property is sometimes referred
to as `pseudo-Hermitian' \cite{Hull:2004in}. However, in the mathematical
literature pseudo-Hermitian usually refers to complex structures which
are isometries of indefinite metrics.} 
Moreover, if $N_{ij}$ is a Hessian metric on 
${\cal S}$, then $K(X,\bar{X}) = 4 {\cal V}(\sigma)$ is a para-K\"ahler
potential and ${\cal E}$ is a para-K\"ahler manifold. Like ${\cal M}$
we can interprete ${\cal E}$ as the tangent bundle of ${\cal S}$, but
equipped with a different metric.

Since the line element of ${\cal E}$ and the corresponding Euclidean
action are indefinite, Derrick's theorem does not apply and it is
possible to find solutions of the equations of motion
\begin{eqnarray}
\partial^m (N_{ij} \partial_m \sigma^j) - \frac{1}{2}
\partial_i N_{jk} (\partial_m \sigma^i \partial^m \sigma^j -
\partial_m \beta^i \partial^m \beta^j) &=& 0\;, \nonumber \\
\partial^m (N_{ij} \partial_m \beta^j) &=& 0 \;.\nonumber
\end{eqnarray}
A particular class, which we call extremal instanton solutions, is
obtained by imposing the ansatz
\begin{equation}
\label{ExtInstAnsatz}
\partial_m \sigma^i = \pm \partial_m \beta^i \;.
\end{equation}
We will see later that this ansatz is indeed related to the saturation
of a Bogomol'nyi bound, and that it is satisfied by $\frac{1}{2}$ 
BPS solutions for Euclidean vector multiplets. Moreover, it implies
that extremal instanton solutions remain unmodified when coupling the
sigma model to gravity, and that one can obtain extremal black hole
solution by dimensional lifting with respect to time. For the time 
being we observe that the extremal instanton ansatz leads to an enormous
simplification of the equations of motion, which now reduce to 
\[
\partial^m (N_{ij} \partial_m \sigma^j) = 0 \;. 
\]
This simplification can be understood in terms of the para-complex
geometry of the scalar target space ${\cal E}$. Geometrically, the
equations of motion for a non-linear sigma model are the equations
for a harmonic map from space-time into the scalar target space.
One way to obtain harmonic maps is to find harmonic maps into
completely geodesic submanifolds 
\cite{Breitenlohner:1987dg,Stelle:1998xg,EucI}. 
The extremal instanton ansatz
(\ref{ExtInstAnsatz}) imposes that tangent vectors to solutions
are null vectors and restricts the solution to take values in a 
totally isotropic submanifold. Totally isotropic submanifolds are
known to be intimately related to extremal solutions, but it remains
to show that the submanifolds defined by (\ref{ExtInstAnsatz}) 
are totally geodesic. But this is automatic if we assume that 
${\cal S}$ is Hessian, so that ${\cal E}$ is para-K\"ahler \cite{EucI}. 
The extremal
instanton ansatz implies that the tangent vectors of the solution
lie within the eigenspaces of the para-complex structure, and for
a para-K\"ahler manifold the para-complex structure is parallel.
Therefore the corresponding submanifolds are totally geodesic. 
We can in fact obtain the solution explicitly. Using that $N_{ij}$
has a Hesse potential ${\cal V}(\sigma)$ we can define
dual scalar fields 
\begin{equation}
\label{DualScalars}
\sigma_i = \partial_i {\cal V} \;,
\end{equation}
and the equations of motion reduce to harmonic equations
\[
\Delta \sigma_i = 0  \;.
\]
Therefore the solution depends on $n$ harmonic functions $H_i(x)$. 
Note that while $\{ \sigma^i, b^i\} $ are uniquely determined 
(up to integration constants for the $b^i$ to be discussed later)
in terms
of the harmonic functions, we can not always solve (\ref{DualScalars})
explicitly for the original scalars $\sigma^i$. The
equations (\ref{DualScalars}) are a set 
of coupled algebraic equations, which, by dimensional lifting over time,
are related to the five-dimensional black hole attractor equations 
\cite{Mohaupt:2009iq}.
We note that the this type of solution relies on the indefinite 
signature of the scalar manifold. For definite signature 
we would only have found constant maps, i.e. constant scalar
fields corresponding to ground state solutions rather than 
instanton solutions.

We also observe that if we substitute the instanton solution $\{\sigma^i_*, 
\beta^i_*\}$
back into the action, we obtain zero
\[
\tilde{S}_E[\sigma_*, \beta_*] = 0 \;.
\]
This is in fact an automatic consequence of the ansatz (\ref{ExtInstAnsatz}),
and thus a flip side of precisely the feature which allows the
existence of non-trivial saddle points in the first place. 
We will see later that a finite
and positive instanton action is obtained by adding a boundary term.

\subsection{Complex saddle points}

The definite and the indefinite Euclidean action are related
by analytical continuation of the axions. Introducing complex
axion fields
\[
B^i = b^i + i \beta^i \;,
\]
the two actions can be obtained as two different `real forms' 
of the complex action
\begin{equation}
\label{EucActComplex}
S_E[\sigma,B] = \frac{1}{2} \int_E d^4 x  N_{ij}(\sigma) 
(\partial_m \sigma^i \partial^m \sigma^j 
+ \partial_m B^i \partial^m B^j) \;.
\end{equation}
The scalar target space parametrized $\{\sigma^i, B^i = b^i + i \beta^i\}$
can be interpreted as the complexified tangent bundle 
$T{\cal S}_{\mathbbm{C}}$ of ${\cal S}$. Note that 
$ds^2 = N_{ij}(\sigma) d B^i dB^j$ is a complex bilinear form
(and not a Hermitian sesquilinear form) on the tangent spaces
$T_P{\cal S}$. 

It is clear that real saddle points of the indefinite
action (\ref{EucActIndef}) can be interpreted as complex (purely imaginary)
saddle points of the definite action (\ref{EucActDef}). Thus the first two of
the ways around Derrick's Theorem mentioned in the introduction, 
complex saddle points and indefinite
actions, are equivalent for this class of models. However, this does
not address the problem that the instanton action is identically
zero, and since we are lead to considering complex field values 
the issue of the damping of the functional integral also deserves
further investigation.

\section{Instanton amplitudes}

In the context of quantum physics, we are interested in instantons
because non-trivial saddle points of the Euclidean functional
integral give rise to non-perturbative corrections. To see 
whether the solutions found above can play such a role, we need
to consider suitable physical quantities, the two obvious candidates
being the partition function and transition amplitudes between 
states of different axionic charge.\footnote{As we will see axionic
charges play a role analogous to instanton charges in Yang-Mills theories,
and therefore we will refer to them sometimes as instanton charges.}

We find it helpful to exploit the analogy with the saddle point 
approximation of one-dimensional real integrals, which often uses
complex variables and contour deformation. As we will demonstrate 
the `complex viewpoint' leads to a transparent and unified treatment
of instantons for axionic scalars, their dual description in terms
of tensor fields, and their lifting to solitonic solutions, including
black holes.

\subsection{A toy example for the saddle point approximation}

The problem of finding a consistent saddle point approximation 
can be illustrated with a simple, one-dimensional toy example. 
We consider an integral over the real line
\[
I = \int_{-\infty}^\infty dx e^{-f(x)}  \;,
\]
which is meant to serve as a toy example either 
for the `axionic' partition function 
\[
Z = \int Db e^{-S_E[b]}  \;,
\]
or for any amplitude that we might want to compute.\footnote{We suppress 
the scalars $\sigma^i$ since the issue we are interested in is the
treatment of the axions.}
Let us assume that $f(x)$ can be continued analytically into 
the complex $z$ plane, $z=x+iy$, 
and that it has a sharp  saddle point at $z_* = i \alpha$, where 
$\alpha$ is real. Then  the
saddle point approximation is obtained by performing a Gaussian 
integration over the saddle point. For concreteness (and since it
will fit with the functional integrals we are interested in) we further
assume that $f(x)$ has a minimum if we pass through the stationary point 
on a contour parallel to the real axis:
\begin{equation}
\label{Damp1}
\partial^2_x f(z)_{z=z_*} >  0 \;.
\end{equation}
The saddle point approximation is obtained by expanding $f$ to second order
and taking the integration contour to be $z(x) = x + i \alpha$, where
$-\infty < x < \infty$:
\begin{eqnarray}
I &\approx& \int_{-\infty + i \alpha}^{\infty + i \alpha} d z 
e^{- f(z_*) - \frac{1}{2} f''(z_*) (z-z_*)^2 } 
= \int_{-\infty}^\infty dx e^{-f(i\alpha) - \frac{1}{2}  \partial^2_xf(i \alpha) x^2}
\nonumber \\
&\simeq& ({\partial^2_x f (i\alpha)})^{-1/2} e^{-f(i\alpha)} \;. \nonumber
\end{eqnarray}
In the analogy, $f(i \alpha)$ is the `instanton action', which should
be positive, finite and non-vanishing, $\partial^2_x  f(i\alpha) >0$
is the `fluctuation determinant', which needs to be positive definite 
so that the Gaussian integral is damped and the saddle point approximation
is consistent. In the following we will verify that the toy example
indeed captures the essential properties of describing instantons 
from the viewpoint of the definite Euclidean action.

To obtain an analogue for the situation concerning the indefinite 
Euclidean action, we compute the same integral using a new,
rotated complex variable $w=-iz$. Note that this is the correct
analogy, because we are not interested in modifying the physical,
Minkowski signature theory, but just use a different Euclidean
continuation to compute the same physical quantity. 
In terms of the rotated variable the function 
$g(w) = f(z(w))$ now has
a real saddle point at $w_* = \alpha$, and since fluctuations 
of $f$ around $z_*$ were damped along the real axis, fluctuations
of $g$ around $w_*$ are damped along the imaginary axis, i.e. 
(\ref{Damp1}) is equivalent to 
\[
\partial^2_v g(w_*) > 0 \;,
\]
where $w=u+iv$. In terms of the rotated variable the integral $I$ 
takes the form
\[
I =-i \int_{i\infty}^{-i\infty} dw e^{-g(w)} 
\] 
and the saddle point approximation takes the form
\begin{eqnarray}
I &\approx& 
-i \int_{i\infty + \alpha}^{-i \infty + \alpha} d w 
e^{- g(w_*) - \frac{1}{2} g''(w_*) (w-w_*)^2 } 
= \int_{-\infty}^\infty dv  e^{-g(\alpha) 
- \frac{1}{2}  \partial^2_vg(\alpha) v^2} \nonumber \\
&\simeq& ({\partial^2_v g (\alpha)})^{-1/2} e^{-g(\alpha)} \;. \nonumber
\end{eqnarray}
We will now demonstrate that this is
the correct analogy for the use of the two Euclidean actions 
for the computation of instanton amplitudes.



\subsection{Instantons from the Wick rotated scalar action}

In this section we 
adapt a classical computation done in \cite{Coleman:1989zu}
for axionic wormholes to our type of models. We closely follow
the presentation given by \cite{Chiodaroli:2009cz} for hypermultiplet
instantons.  The amplitude to be computed is the 
transition amplitude between
two different field configurations of the axions, which are 
specified on spatial hyperplanes. 
We will treat the $\sigma^i$ as
spectators in the following, and suppress them in most of the 
formulae.
Let us denote the initial and final field configurations 
of the fields $b^i(x)$ by $\chi^i_I({\bf x})$ and $\chi^i_F({\bf x})$,
respectively. Here and in the following $x$ are coordinates on
Euclidean space $E= \mathbbm{R}_t \times \mathbbm{R}^3$, $t$ is Euclidean
time and ${\bf x}$ are coordinates on space $\mathbbm{R}^3$. 
The asymptotic  configurations are imposed at the times
$t_I$ and $t_F$, which we ultimately take to the limits
$t_I \rightarrow -\infty$ and $t_F \rightarrow \infty$. 

We are interested in computing the quantum transition amplitude
between the initial state $|I\rangle = |\chi_I \rangle$ and the
final state $|F\rangle =| \chi_F \rangle$. Here 
$|I\rangle = |\chi_I\rangle =
|\chi^i_I({\bf x}) \rangle$ is a formal `position' eigenstate, where
the fields $b^i$ assume the configuration $\chi^i_I({\bf x})$. 
The corresponding  Euclidean transition amplitude between the two states
is
\[
{\cal A} = \langle F | e^{-H(t_F - t_I)} |I \rangle \;,
\]
where $H$ is the Hamiltonian. 
In the functional formulation this becomes
\[
{\cal A} = \int_{BC} D b e^{-S_E[b]} \;,
\]
where the fields $b^i(x)$ are subject to the boundary conditions
$b^i({\bf x},t_I) = \chi^i_I({\bf x})$ and 
$b^i({\bf x},t_F) = \chi^i_F({\bf x})$.
As announced, we suppress the fields $\sigma^i$ for the time being.

As usual in quantum mechanics and quantum field theory, we are 
interested in computing transition amplitudes
for asymptotically large time. The resulting asymptotic transition 
amplitudes are time-independent, and there is no need to explicitly
continue back the result from Euclidean to physical time, in contrast
to Greens functions. In the limit of asympotically large time only the
lowest eigenvalue of the Hamiltonian contributes. While naively this
would restrict one to compute vacuum amplitudes, one can also access
amplitudes between the ground states of charge (superselection) sectors
by inserting suitable projection operators \cite{Coleman:1989zu}. In our case 
the theory has $n$ conserved charges related to the $n$ 
commuting shift symmetries, and we can ask for transition 
amplitudes between states with 
fixed initial and final axionic charges $Q_i^{I/F}$.
Since states are characterized by specifying field configurations
rather than global (integrated) quantities, the relevant 
quantity to compute is the transition amplitude
between states with prescribed 
initial and finite charge densities $\rho_i^{I/F}({\bf x})$,
where $Q_i^{I/F} = \int_{I/F} d^3 {\bf x} \rho_i^{I/F} ({\bf x})$. 
The projection of the amplitude ${\cal A}$ 
onto such states amounts to the insertion 
of a functional delta function into the amplitude, which 
forces the time-like components of the  
Noether currents $j_{m|i}(x)$ to take prescribed values at
$t_I$ and $t_F$. Specifically, for the initial state we have
\[
P_I |I \rangle = 
\delta( \rho^{I} - j_{t}^{I}) | \chi_I \rangle \;,
\]
where $j_{t}^I = j^I_{t|i}({\bf x}) = j_{t|i}({\bf x},t_I)$ is the
Noether current at time $t=t_I$, and the delta function 
is a formal `functional delta function'
\[
\delta( \rho^{I} - j_{t}^{I}) = \prod_{{\bf x}\in \mathbbm{R}^3_{t=t_I}}
\prod_{i=1}^n \delta( \rho_i^{I}({\bf x}) - j_{t|i}^{I}({\bf x})) \;,
\]
which imposes the boundary values for the Noether current.

In order to implement this projection in the functional integral,
we use the Fourier representation of the delta function. For a
functional delta function this leads to an additional functional
integral over auxiliary functions $\gamma_I=(\gamma_I^i({\bf x}))$ 
and $\gamma_F = (\gamma_F^i({\bf x}))$ which 
`live' on the initial and final hypersurface:
\[
P_I |I \rangle = \int_I D \gamma_I 
e^{-i \int_I d^3 {\bf x} (\rho^I - j_{t}^{I})\cdot
\gamma_I} |\chi_I \rangle \;.
\]
To avoid cluttered formulae we use the letters $I$ and $F$ to indicate
that an integral is taken over the initial or final hypersurface,
or that a functional integral is taken over functions on the initial
and final hypersurface. The arguments $x$, ${\bf x}$ are omitted 
where possible without ambiguity. We use the short hand notation
\[
(\rho^I - j_{t}^{I})\cdot \gamma_I = 
(\rho^I_i({\bf x}) - j_{t|i}^I({\bf x})) \gamma_I^i({\bf x}) \;.
\] 
We now use that the Noether charges $Q_i$ generate shift symmetries 
$b^i \rightarrow b^i + C^i$. Since $j_{m|i}$ is the corresponding
Noether current, the combination  $j_t^I \cdot \gamma_I$ 
generates shifts with (local) parameter $\gamma_I^i({\bf x})$,
and acts on states as $|\chi_I\rangle \rightarrow |\chi_I + \gamma_I\rangle$. 
Therefore
\[
P_I | I \rangle =
\int_I D \gamma_I e^{-i \int_I d^3 {\bf x} \rho^I \cdot \gamma_I}
|\chi_I + \gamma_I \rangle
\]
and similarly 
\[
P_F | F \rangle =
\int_F D \gamma_F e^{-i \int_F d^3 {\bf x} \rho^F \cdot \gamma_F}
|\chi_F + \gamma_F \rangle \;.
\]
The charge density projected amplitude is
\begin{eqnarray}
&&\tilde{\cal A} = \langle F | P_F e^{-H(t_F - t_I) }P_I | I \rangle
\nonumber \\
&=& \int_F D \gamma_F \int_I D \gamma_I
\int_{BC} D b
e^{i \int_F d^3 {\bf x} \rho^F \cdot \gamma_F }
e^{-i \int_I d^3 {\bf x} \rho^I \cdot \gamma_I}
e^{- S_E[b] } \;,
\end{eqnarray}
with boundary conditions 
\[
b^i({\bf x},t_I) = \chi^i_I({\bf x}) + \gamma^i_I({\bf x}) \;,\;\;\;
b^i({\bf x},t_F) = \chi^i_F({\bf x}) + \gamma^i_F({\bf x})\;.
\]
The three functional integrals can be 
combined into a single integral over 
fields $b^i(x)$ without boundary conditions.
Setting
\[
\gamma_{I/F} = \tilde{\gamma}_{I/F} - \chi_{I/F} \;,
\]
the amplitude becomes
\[
\tilde{\cal A}= 
\int_F D \tilde{\gamma}_F \int_I D \tilde{\gamma}_I
\int_{BC} D b
e^{i \int_F d^3 {\bf x} (\tilde{\gamma}_F-\chi_F) \cdot \rho^F }
e^{-i \int_I d^3 {\bf x}(\tilde{\gamma}_I - \chi_I) \cdot \rho^I }
e^{- S_E[b] } \;,
\]
with boundary conditions
\[
b^i({\bf x},t_{I/F}) = \tilde{\gamma}^i_{I/F}({\bf x}) \;.
\]
The functional integrals over $\tilde{\gamma}_{I/F}$ are now
integrations over boundary conditions for the $b^i(x)$, so that
we obtain an integral over $b^i(x)$ without boundary
conditions,
\[
\tilde{\cal A} = e^{-i\int_F \chi_F \cdot \rho^F + i\int_I \chi_I \cdot \rho^I} 
\int D b e^{i \int_F b^i \rho_i^F - i \int_I b^i \rho_i^I} e^{-S_E[b]}  \;.
\]
The phase factor in front of the remaining functional integral
depends on the field configurations
$\chi_{I/F}$ and charge densities $\rho^{I/F}$ which define the
initial and final states $P_{I/F} |I/F\rangle$. This pre-factor 
keeps track of the relation between the `position eigenstates'
and the `charge eigenstates'.
The remaining functional integral is a transition amplitude between
charge (density) eigenstates. While the integration over the fields
$b^i$ is now unrestricted, this integral is still sensitive to the 
boundary conditions, which encode the physical states, through the
boundary term. As we will see this boundary term determines the saddle
point which dominates the saddle point approximation. If the boundary
term is dropped, the remaining functional integral is simply the partition 
function. As we will see the same boundary term also occurs in various
other contexts (dimensional lifting and Hodge dualization), so that
it is natural to regard $S_E + \Sigma$ as a modified or `improved'
action. For the amplitude at hand the boundaries which support the
boundary term are spatial hyperplanes located at $t=t_I$ and $t=t_F$,
respectively. The two boundary terms can be combined into the more
compact expression 
\[
\Sigma = i \int_I d^3 {\bf x} b^i \rho_i^I - i \int_F d^3{\bf x}
b^i \rho_i^F = - i \oint_{\partial E} b^i \rho_i  \;,
\]
where $\partial E$ is the combined boundary, with the orientation chosen 
such that the future directed normal is the outer normal.\footnote{Thus
in terms of homology classes, $[\partial E] = [F] - [I]$.}
We remark in passing that the last formula can be used when 
other types of boundaries are relevant. For example, for spherically
symmetric instanton solutions, such as hypermultiplet or vector multiplet
instantons, 
the boundary is an asymptotic three sphere,
$\partial E = S^3_{(\infty)}$. As we will see later such instanton solutions
can be lifted to black hole solutions.

Let us summarize our result for the amplitude
\[
\tilde{\cal A} = \tilde{\cal A}(\chi_I, \chi_F, \rho^I, \rho^F) 
= e^{-i\oint \chi \cdot \rho} \int D b 
e^{-S_E[b] - \Sigma[b|\rho]}  \;,
\]
before proceeding to the 
saddle point approximation.
When identifying critical points of the action, we have to take
into account the boundary term. We consider `boundary variations'
and `bulk variations' separately, starting with the boundary 
variations. Here two contributions arise. The obvious one is
the variation of the boundary term:
\[
\delta \Sigma = - i \oint d^3 {\bf x} \delta b^i \rho_i \;.
\]
But boundary terms also arise from variations of the bulk action 
upon integration by parts.
\begin{eqnarray}
\delta S &=& \int d^4 x N_{ij}(\sigma) \partial_m \delta b^i \partial^m  b^j 
\nonumber \\
&=& \int d^4 x \partial_m (N_{ij}(\sigma) \partial^m b^j \delta b^i) 
- \int d^4 x \partial_m (N_{ij}(\sigma) \partial^m b^j) \delta b^i
\nonumber \\
&=& \oint d^3 {\bf x} n^m N_{ij}(\sigma) \partial_m b^j \delta b^i  
- \int d^4 x \partial_m (N_{ij}(\sigma) \partial^m b^j) \delta b^i \;.
\end{eqnarray}
Here $n^m$ is the outer unit normal vector of the boundary. Note 
that this vector points into the positive direction on the final
hypersurface and into the negative direction on the initial hypersurface.
The second term gives rise to the equation of motion 
of the $b^i$,
\[
\partial_m (N_{ij}(\sigma) \partial^m b^j) = 0 \;,
\]
which expresses conservation of the current associated with the 
shift symmetry.  The first term contributes to the variation 
at the boundary, and the resulting total boundary variation is
\begin{eqnarray}
(\delta S + \delta \Sigma)_{\rm boundary} &=&
\oint (  n^m N_{ij}(\sigma) \partial_m b^j \delta b^i  
- i \delta b^i \rho_i )  \\
&=& \int_F d^3 {\bf x} 
( N_{ij}{\sigma} \partial_t b^j - i \rho_i^F) \delta b^i 
- \int_I d^3{\bf x} 
(N_{ij}{\sigma} \partial_t b^j - i \rho_i^I) \delta b^i \;.
\nonumber
\end{eqnarray}
Vanishing of the boundary variation implies
\begin{eqnarray}
(N_{ij}(\sigma) \partial_t b^j)_{t=t_I}   &=&  i \rho_i^I({\bf x}) \;,
\nonumber \\
(N_{ij}(\sigma) \partial_t b^j)_{t=t_F}   &=&  i \rho_i^F({\bf x}) \;.
\end{eqnarray}
The factors $i$ might at first appear surprising. The functions 
$\rho_i^{I/F}({\bf x})$ are the charge densities of the physical
states entering our amplitude and must therefore be real, and thus
the saddle point solution for the $b^i$ must satisfy 
imaginary  boundary conditions. While this might be counter-intuitive,
these boundary conditions are
physically meaningful, because, following \cite{Coleman:1989zu}, we have
projected onto eigenstates of the Euclidean Noether current. The
Wick rotation introduces an a factor of $i$ in the time component
of vectors,
and if we write the boundary condition in terms of Minkowski time, we 
obtain
\[
(N_{ij}(\sigma) \partial_0 b^j)_{I/F} = \rho_i^{I/F} \;.
\]
Moreover, inspection of the bulk variation shows that the
saddle point solution which dominates the Euclidean functional 
integral is indeed imaginary.  

In order to find explicit and non-trivial saddle points we need to 
re-instate the scalar fields $\sigma^i$, which we have neglected so far.
Then the  full bulk action is 
\begin{equation}
\label{EucActBulk}
S_E[\sigma,b] = \frac{1}{2} \int d^4 x N_{ij}(\sigma) (\partial_m \sigma^i 
\partial^m \sigma^j + \partial_m b^i \partial^m b^j) \;.
\end{equation}
The bulk variation of $b^i$ is not modified. 
The variation of the $\sigma^i$ gives their equation of motion 
\[
\partial^m (N_{ij}(\sigma) \partial_m \sigma^j) - \frac{1}{2}
\partial_i N_{jk}(\sigma) (\partial_m \sigma^j \partial^m \sigma^k
+ \partial_m b^j \partial^m b^k) = 0\;.
\]
The extremal instanton solutions discussed previously are
obtained by imposing the extremal instanton ansatz
\begin{equation}
\partial_m \sigma^i = \pm i \partial_m b^i \;.
\end{equation}
Since we are working with the definite Euclidean action, the
saddle point solution of the bulk action  $b^i_*$ is imaginary. 
This is consistent with the imaginary boundary conditions 
which we found from the boundary variation, and therefore the
saddle point  contributes to the amplitude we are computing.

Let us therefore turn to the saddle point evaluation of
the amplitude $\tilde{\cal A}$. We denote the saddle point 
solution by $\sigma^i_*$ and $b^i_* = i \beta^i_*$, where 
$\beta^i_*$ is real and satisfies the boundary condtions
\[
(N_{ij}(\sigma) \partial_t \beta^j)_{*|I/F} = \rho_{i}^{I/F} \;.
\]
Since we have only specified initial and final states of
$b^i$, and not of $\sigma^i$, there are no boundary conditions
for the $\sigma^i$. We only impose that they satisfy their bulk
equations of motion, and this will lead to a
consistent semi-classical transition amplitude for the $b^i$.
Thus the $\sigma^i$ are treated as classical on shell
background fields.

The leading contribution to the quantum amplitude $\tilde{A}$
comes from evaluating the action at the saddle point. 
As we have already noted previously, the 
bulk action vanishes at the saddle point:
\[
S_E [\sigma_*, b_*] = \frac{1}{2}  d^4 x \int N_{ij}(\sigma_*) 
(\partial_m \sigma^i_* \partial^m \sigma^j_* 
+ \partial_m b^i_* \partial^m b^j_*)  = 0 \;,
\]
as a consequence of the
ansatz (\ref{ExtInstAnsatz}). However we now see that
the day is saved by the boundary action, which gives
\begin{equation}
\label{InstAct_b}
\Sigma[b_*] = - i \oint d^3 {\bf x} b^i_* \rho_i = \oint d^3 {\bf x}
\beta^i_* \rho_i 
= \beta^i_*(t_F) Q_i^F - \beta^i_*(t_I) Q_i^I \;,
\end{equation}
and therefore
\[
\tilde{\cal A} \propto e^{-S_* - \Sigma_*} = 
e^{-\Sigma_*} = e^{-\beta^i(t_F) Q_i^F - 
\beta^i(t_I) Q_i^I} \;.
\]

In particular, if $\beta^i_*(t_F)=\beta^i_*(t_I)=\beta_0$, then
$\Sigma_* = \beta_0 ( Q_i^F - Q_i^I)$. Thus the instanton
amplitude is proportional to the difference between the charges
of the initial and final state, as expected for a tunneling 
amplitude between ground states of charge (superselection) sectors.
We remark that in contrast to the classical action the transition 
amplitude depends 
on the values  $\beta^I_*(t_{F/I})$ of the axions. 
The classical continuous shift
symmetry is broken to a discrete subset of imaginary shifts 
in $\beta^i$ and thus real shifts in $b^i$:
\[
\beta^i \rightarrow \beta^i + 2 \pi i k \Leftrightarrow
b^i \rightarrow b^i + 2 \pi k \;,\;\;\;k \in \mathbbm{Z} \;.
\]
The breaking of continuous shift symmetries by instantons
is a typical quantum effect. Since the shift symmetry is a global
symmetry it does not lead to an inconsistency of the theory.

Finally, a consistent saddle point approximation requires that 
the functional integral is damped, i.e. that the fluctuation 
determinant is positive definite. 
This is
indeed the case, as long as the functional integration is performed
along `real directions' in $b^i$-field space. More precisely, 
when shifting the integration variable according to 
\begin{equation}
\label{Saddle+Fluct}
b^i = i \beta_*^i + \tilde{b}^i \;,
\end{equation}
where $\beta_*^i$ is the saddle point solution, and where the 
(real) fluctuation 
$\tilde{b}^i$ is the new integration variable, then
\[
\tilde{\cal A} \propto e^{-\Sigma_*} \int D\tilde{b} 
e^{- \frac{1}{2} \int d^4 x N_{ij}(\sigma_*) \partial_m \tilde{b}^i 
\partial^m \tilde{b}^j } \;.
\]
Since $N_{ij}$ is assumed to 
be positive
definite, the fluctuation determinant is damped and leads to a
Gaussian integral. Therefore the saddle point approximation 
is well defined. 
Comparing (\ref{Saddle+Fluct}) to the one-dimensional toy example
we see that this indeed captures the essential features of the
saddle point approximation of the axionic functional integral.

\subsection{Instantons from the indefinite scalar Euclidean action}

Let us now address the issue whether the indefinite action (\ref{EucActIndef})
can play a role in computing instanton amplitudes, despite that it is not
positive definite. Here the key point is to note that irrespective of
how we define an (or `the') Euclidean version of the theory, we do not
intend to change the physical, Minkowski signature theory. Therefore 
we still want to
compute the same physical amplitude as before, albeit using a different
analytical continuation.
The initial and final states of the amplitude 
are `position' or `charge' eigenstates
of the original Minkowski signature field $b^i$. Thus when using 
the rotated axion field $\beta^i = - i b^i$, the boundary condition 
of the functional integral which encodes the initial and final 
state are
\begin{equation}
\label{BCbeta}
i \beta^i({\bf x}, t_{I/F}) = \chi^i_{I/F}({\bf x}) \;.
\end{equation}
While within our approach 
it is clear from the start that we are just rewriting the amplitude 
in terms of a different, rotated integration variable, it is
instructive to revisit some of the key formulae of the previous
section and to express them in terms of the variable $\beta^i$. 
We already saw that the variation of the bulk action $\tilde{S}_E$ with 
respect to the rotated axions 
$\beta^i$ leads to a real saddle point $\beta^i= \beta^i_*$. 
However the boundary conditions (\ref{BCbeta}) of the functional integral over 
$\beta^i$  
are now imaginary.\footnote{We are referring to the boundary conditions
of the functional integral representing the amplitude, not the boundary
conditions satisfied by the saddle point solution.} 
The charge projected amplitude can be brought to  the form
\[
\tilde{A} = e^{-i \oint \chi \cdot \rho} \int D \beta 
e^{-\tilde{S}_E[\beta] - \tilde{\Sigma}[\beta| \rho]} \;.
\]
The prefactor is the same as before, since $\chi$ and $\rho$ 
characterize physical position and charge eigenstates and are
not subject to analytical continuation. 
In the saddle point approximation the $\beta^i$ decompose
into a real saddle point solution $\beta^i_*$ and 
purely imaginary fluctuations $\tilde{\beta}^i = - i \tilde{b}^i$:
\[
\beta^i = \beta^i_* + \tilde{\beta}^i = \beta^i_* - i \tilde{b}^i \;.
\]
As in the rotated version of the one-dimensional toy example the
integration, the original integration is over purely imaginary 
field configurations (`integration along the imaginary axis'), and 
get
shifted by the real saddle point solution.

The boundary term is
\[
\tilde{\Sigma} = \oint d^3{\bf x} \beta^i \rho_i 
\]
and the vanishing of the total boundary variation (variation of
boundary terms plus terms obtained by integration by parts)
implies
\[
(N_{ij}(\sigma) \partial_t \beta^j)_{I/F} = \rho_i^{I/F}
\]
which is consistent with a real saddle point solution. 
Also note that the Euclidean Noether current for the indefinite
action is indeed
real, because both time and the axion field have been 
Wick rotated. 

The integration over fluctuations takes the form
\[
\int D\tilde{\beta} e^{\frac{1}{2} \int d^4 x N_{ij} \partial_m \tilde{\beta}^i
\partial^i \tilde{\beta}^j } \;.
\]
This is damped for positive definite $N_{ij}$ 
if the fluctuations are purely imaginary. This 
is consistent with the imaginary boundary conditions of the 
original integral that we approximate.  

So far we have insisted that the two Euclidean actions we compare
are Euclidean versions of the same Minkowski signature theory. 
Alternatively, one might take the viewpoint that the Euclidean 
theory is to be taken fundamental. In this approach a theory
is defined by its Euclidean functional integral, and physical
quantities in Minkowski space are obtained by analytical continuation
in either position or momentum space. From this point of view 
it is conceivable that physically inequivalent Euclidean actions
could be found. But for the class of actions considered here 
it is clear as a result of the above analysis that  
the requirement of a consistent saddle point approximation does 
only admit one consistent choice, modulo a choice of variables.
We remark that the situation seems to become more complicated if 
sigma models with a more complicated, non-Abelian isometry group
are considered. In particular, as observed in \cite{Chiodaroli:2009cz},
it can happen that some of the saddle points of the combined bulk and 
boundary action do not have a real positive instanton action.
These complications are also reflected by the observation that
the universal hypermultiplet admits 
that there exist three different real Euclidean versions, one with 
a definite and
two with an indefinite target space, 
which are different real sections of one underlying
complex action \cite{Theis:2002er,Gunaydin:2007bg,Mohaupt:2008zz}.
Thus once the isometry group is not required to be Abelian,
the question which complex saddle points contribute in the semi-classical
approximation becomes more complicated, and there are more 
Euclidean actions to be considered. We plan to investigate
actions with non-Abelian isometries in the future, based on generalized
version of the c-map. 

\section{Instantons from the Euclidean scalar-tensor action}

We now turn to the third Euclidean formulation of the theory,
where the axionic scalars $b^i$ are replaced by antisymmetric
tensor fields $B_{mn|i}$. Aspects of this dualization for the
class of models under consideration were discussed in detail 
in \cite{Mohaupt:2009iq}. Therefore we start with a brief summary of
the points relevant for the present discussion, and then add 
further remarks concerning the relation between the partition functions
and the role of axionic zero modes.

The field strength of the antisymmetric tensor fields are
\[
H_{mnp|i} = 3! \partial_{[m} B_{np]|i} \propto
N_{ij}(\sigma) \epsilon_{mnpq} \partial^q b^j \;.
\]
The Euclidean action of the scalar-tensor theory 
\begin{equation}
\label{EucActB}
S_E[\sigma,B] = \int d^4 x \left( 
\frac{1}{2} N_{ij}(\sigma) \partial_m \sigma^i
\partial^m \sigma^j + \frac{1}{2 \cdot 3!} N^{ij}(\sigma) H_{mnp|i}
H^{mnp}_j \right) \;,
\end{equation}
is obtained from its Minkowski counter part by the standard 
Wick rotation.
This Euclidean action is positive definite, and a 
Bogomol'nyi bound
is found by re-writing it as a sum of squares, plus a
remainder:
\[
S_E[\sigma,B] = \int d^4 x \left[ \frac{1}{2} \left(
\partial_m \sigma^i \mp \frac{1}{3!} N^{ij}(\sigma) \epsilon_{mnpq}
H^{npq}_j \right)^2 \pm \frac{1}{3!} \partial_m \sigma^i 
\epsilon^{mnpq} H_{npq|i} \right] \;.
\]
Thus the action is minimized by imposing
\begin{equation}
\label{DualInstAnsatz}
\partial_m \sigma^i = \pm \frac{1}{3!} N^{ij}(\sigma) 
\epsilon_{mnpq} H^{npq}_j \;,
\end{equation}
which is the Hodge-dual version of the 
extremal instanton ansatz (\ref{ExtInstAnsatz}). 
After imposing this ansatz, the remaining scalar equations 
of motion for the $\sigma^i$ are the same as in the scalar
formulation of the theory, and can be solved in terms of
harmonic functions by passing to dual scalars. Substituting
the solution back into the action, one obtains
\[
S_{\rm inst} = \left. \int d^4 x N_{ij}(\sigma) \partial_m \sigma^i 
\partial^m \sigma^j \right|_*\;,
\]
where the $\sigma^i_*$ are obtained by solving (\ref{DualScalars})
in terms of harmonic functions. While resulting from the bulk 
action (\ref{EucActB}) this  is a 
boundary term, modulo the equations of motion:
\[
S_{\rm inst} = \left. \oint d^3 {\bf x} n^m N_{ij}(\sigma) 
\sigma^i \partial_m \sigma^j \right|_* \;.
\]
The magnetic charges with
respect to the $B$-fields are
\[
Q_i = \frac{1}{3!} \oint d^3 {\bf x} n^m \epsilon_{mnpq} H^{npq}_i \;,
\]
where we choose the normalization 
such that they are 
equal to the electric charges with respect to the $b$-fields.
When evaluated on instanton solutions we can use (\ref{DualInstAnsatz})
to rewrite this expression as
\[
Q_i = \left. \oint d^3 {\bf x} n^m N_{ij}(\sigma) \partial_m \sigma^j 
\right|_* \;.
\]
The resulting instanton action is 
\begin{equation}
\label{InstAct_sigma}
S_{\rm inst} = \left. \sigma^i(t_F) Q_i^F - \sigma^i(t_I) Q_i^I \right|_*\;,
\end{equation}
for solutions which interpolate between boundary conditions
imposed on hypersurfaces located at Euclidean times $t_F$ and $t_I$.
If we compare this to the instanton action (\ref{InstAct_b}) 
found when computing the
instanton amplitude using the two axionic actions, 
\[
\Sigma_* = \left. \beta^i(t_F) Q_i^F - \beta^i(t_I) Q_i^I \right|_*\;,
\]
we find that the results almost but not quite agree. The reason 
is that the relation $\partial_m \sigma^i = \pm \partial_m \beta^i$
only fixes the $\beta^i$ up to integration constants. As far
as classical physics is concerned these integration constants are 
irrelevant because of the continuous shift symmetry. However, as we
have seen, this symmetry is broken in the quantum theory to a discrete
subgroup, and the quantum theory is sensitive to the values of the axions
and hence to the integration constants $C^i= \sigma^i(t_F) - \beta^i(t_F)
= \sigma^i(t_I) - \beta^i(t_I)$. Instanton actions based on axions and
on antisymmetric tensor fields differ in general by the amount
\[
S_{\rm inst} - \Sigma_* = C^i (Q_i^F - Q_i^I) \;.
\]
If we insist that the quantum theories based on axions and antisymmetric
tensor fields are equivalent, then we must impose that the $C^i$ vanish
modulo the remaining discrete shift symmetries
\begin{equation}
\label{Cquantisation}
C^i = 2 \pi i k \;,\;\;\;k \in \mathbbm{Z} \;.
\end{equation}
This is a restriction on the zero modes of the axion fields $b^i$,
which classically are completely trivial. The subtlety that we find
in the quantum equivalence between axions and antisymmetric tensor
fields reflects that the axionic shift symmetry is a global symmetry
which is broken generically by quantum effects, while the tensor
field theory has a local gauge symmetry, which cannot. The condition 
(\ref{Cquantisation}) imposes that the charge sectors and saddle points
of both theories match.  We will see
later that the relation (\ref{Cquantisation})
is equivalent to imposing that the instanton
action equals the mass of the soliton obtained by dimensional lifting.

Let us next discuss the relation between 
the Euclidean scalar-tensor action (\ref{EucActB}) 
and the two Euclidean scalar actions (\ref{EucActDef}) and 
(\ref{EucActIndef}) in more detail. At the classical level the
dualization is performed by first adding the Bianchi identies for the
field strength $H_{mnp|i}$ with Lagrange multipliers, denoted $b^i$,
and then eliminating the tensor fields by their equation of motion.
Since details were given in \cite{Mohaupt:2009iq}, we only cite the
final result 
\begin{eqnarray}
S_E[\sigma,b] &=& \int d^4x \left( \frac{1}{2} N_{ij}(\sigma)
\partial_m \sigma^i \partial^m \sigma^j 
- \frac{1}{2}(3!\lambda)^2 N_{ij}(\sigma) \partial_m b^i 
\partial^m b^j \right) \nonumber \\
&& + (3!\lambda)^2 
\oint d^3 {\bf x} n^m b^i N_{ij}(\sigma) \partial_m b^j \;.\nonumber
\end{eqnarray}
Here $\lambda$ is normalization constant. We observe that
either the definite scalar-axion action (\ref{EucActDef}) or the 
indefinite scalar-axion action (\ref{EucActIndef}) is obtained,
with the same boundary term as in the amplitude
calculation, by setting $(3!\lambda)^2= -1$ and $(3!\lambda)^2 =1$,
respectively. The first choice, corresponding to imaginary
$\lambda$ preserves the definiteness of the action, but maps
the real instanton solution to an imaginary saddle point. 
The second choice, corresponding to real $\lambda$, 
preserves the reality of the saddle point, but yields an 
indefinite Euclidean action. This is a particular feature 
of Euclidean signature. In Minkowski signature dualization simultanously
preserves the saddle points and the positivity of kinetic terms.
The fact that in the Euclidean formulations of axions we either
loose definiteness of real saddle points confirmis our previous
remarks that it is unavoidable, and in fact natural, to
consider complex field configurations.

We also observe that the boundary term obtained by dualization 
contains the undifferentiated axions $b^i$. Therefore the resulting
instanton action agrees with the one (\ref{InstAct_b}) 
obtained in the amplitude calculation. As a consequence, dualization 
does not preserve the value of the instanton action, unless we impose
the condition (\ref{Cquantisation}). The same conclusion is reached
if the dualization is performed at the level of the functional
integral instead of the classical action. This could be demonstrated
by repeating the calculation of the transition amplitude between
axionic states. We prefer to consider the dualization of the 
partition function instead, because this shows that the subtleties
related to the axionic zero modes even occur when no boundary 
conditions (corresponding to initial and final states) are imposed
on the functional integral.

Since the scalar fields $\sigma^i$ are spectators in the dualization,
we only consider the part of the partition function which 
depends on the tensor fields. The starting point is a simplified
partition function,\footnote{When working with the $B$-field itself, 
one would need
to include gauge fixing terms, ghosts, and ghosts for ghosts, since this is a
reducible gauge theory, as reviewed in 
\cite{HenneauxTeitelboim,Mohammedi:1995ca}.
But these complications do not appear to be relevant
to our purpose.} where the Bianchi identity is implemented
through a functional delta function, so that we can perform 
an unrestricted integral over the field strenght $H_{mnp|i}$:
\[
Z = \int DH \exp \left( - \int d^4 x \frac{1}{2\cdot 3!} 
N^{ij} H_{mnp|i} H^{mnp}_j \right) \delta(\epsilon^{mnpq} 
\partial_m H_{npq|k})  \;.
\]
The first step is to convert the functional delta function 
into an additional `Fourier'
functional integral:
\[
= \int DH Db \exp\left( - \int d^4x ( \frac{1}{2\cdot 3!} 
N^{ij} H_{mnp|i} H^{mnp}_j - i \mu b^i \epsilon^{mnpq}
\partial_{m} H_{npq|i} ) \right) \;.
\]
Here $\mu$ is a constant that we will use later to obtain the
conventional normalization of the resulting dual action.\footnote{We will
find that $\mu$ is related to the normalization constant occuring 
in the `classical' dualization by $\mu = \pm i \lambda$.}  
Next we perform an integration by parts, so that the $H$-field
only occurs algebraically, or within boundary terms:
\begin{eqnarray}
Z &=& \int DH Db \exp \left(
\int d^4 x ( - \frac{1}{2\cdot 3!} N^{ij} H_{mnp|i} H^{mnp}_j
+ i \mu \partial_m b^i \epsilon^{mnpq} H_{npq|i} ) \right. \nonumber \\
&& \left. - \mu i \int d^4 x \partial_m (\epsilon^{mnpq} b^i H_{npq|i}) 
\right) \;.
\end{eqnarray}
Defining
\[
\tilde{H}_{mnp|i} = H_{mnp|i} - 3! \mu i N_{ij} \epsilon_{mnpq} \partial^q
b^j \;
\]
we can complete the square and shift integration variables 
$H_{mnp|i} \rightarrow \tilde{H}_{mnp|i}$ to obtain
\begin{eqnarray}
Z &=& \int D\tilde{H} D b \exp \left(
- \int d^4x N^{ij} \tilde{H}_{mnp|i} \tilde{H}^{mnp}_j 
- \int d^4 x (3!\mu) N_{ij} \partial_m b^i \partial^m b^j
\right. \nonumber \\
&&
\left. - \mu i \oint n^m b^i \epsilon_{mnpq} \tilde{H}^{npq}_i 
+ (3!\mu)^2 \oint n^m N_{ij} b^i \partial_m b^j \right) \;.
\end{eqnarray}
The integration over $\tilde{H}_{mnp|i}$ decouples from the
integration over $b^i$, except for one of the boundary terms.
This term vanishes for on shell field configurations where 
$H_{mnp|i}$ and $\partial_q b^i$ are Hodge dual, and we
assume that it can be dropped consistently within the 
semiclassical approximation. Given this, the integration over 
$\tilde{H}_{mnp|i}$ decouples and gives a (formal, infinite)
multiplicative constant that we can ignore. The remaining
functional integral for the fields $b^i$ is
\[
Z = \int Db \exp \left( 
- \int d^4 x (3!\mu) N_{ij} \partial_m b^i \partial^m b^j
+ (3!\mu)^2 \oint n^m N_{ij} b^i \partial_m b^j \right) \;,
\]
which is the partition function associated with the definite
Euclidean action (\ref{EucActDef}), 
together with the boundary term. The standard
normalization is obtained by choosing $(3!\mu)^2=1$, which 
implies that $\mu$ and the previous normalization constant 
$\lambda$ are related by $\mu = \pm i \lambda$. Within the
functional framework it is natural to regard $\mu$ as a real
constant, because the $b^i$ are introduced through the 
Fourier representation of the functional delta function 
implementing the Bianchi identity. Then 
the definiteness of the Euclidean action is preserved,
so that the functional integral remains damped. However, 
we cannot avoid completely to consider complex field values:
the shift in the integration over the $H$-fields is imaginary,
and the real saddle points of the tensor action which correspond
to instantons become imaginary saddle points upon dualization.
Thus we need to consider complex values of the axion fields 
$b^i$ to match the semiclassical expansions of both version of the
theory. Moreover, the presence of the boundary term breaks the
continuous shift symmetry of the bulk action once we consider
field configurations which have support on these boundary terms. This 
feature has no counter part in the tensor field partition function.

\section{Relating instantons to  solitons}

To round up our investigations, we now briefly relate them
to the results of \cite{Mohaupt:2009iq}, where the same class of 
instanton solutions was used to generate solitons and black holes
by dimensional lifting. The main point is that instanton action
is directly related to the mass of the solutions obtained by lifting.

\subsection{Lifting without gravity}

As shown in \cite{Mohaupt:2009iq} the indefinite Euclidean action (\ref{EucActIndef})
can be lifted with respect to time to a theory of real scalars
and abelian gauge fields of the form
\begin{equation}
\label{5dActionNonSusy}
S =  \int d^5 x \left(- \frac{1}{2} N_{ij} (\sigma) \partial_\mu
\sigma^i \partial^\mu \sigma^J  - \frac{1}{4} F^i_{\mu \nu}
F^{j|\mu \nu} \right) \;,
\end{equation}
where $\mu, \nu = 0,1,2,3,4$ are five-dimensional Lorentz indices.
The four-dimensional scalars $b^i$ are related to the five-dimensional
gauge fields $A^i_\mu$ by $b^i = - A_0^i$. The scalar action 
(\ref{EucActIndef}) does not account for the magnetic components
$F^i_{mn}$ of the five-dimensional gauge fields. However, (\ref{EucActIndef})
is a consistent truncation in the sense that any solution of (\ref{EucActIndef})
lifts to a static, purely electric solution of (\ref{5dActionNonSusy}).
If one uses (\ref{EucActIndef}) to generate five-dimensional solitons,
one could add further terms to the five-dimensional action as long
as they do not contribute to static, purely electric backgrounds.
One example, which occurs in supersymmetric theories, is a Chern-Simons
term. The lifting works without imposing any restrictions on the 
scalar metric $N_{ij}$, bu to obtain explicit solutions in terms of harmonic
functions we impose that $N_{ij}$ is a Hessian metric. 
Upon reduction over time we obtain para-K\"ahler metric on the 
resulting extended scalar manifold ${\cal E}$. Let us also note
that by dimensional reduction over space we obtain the 
Minkowski signature action (\ref{MinkAct}) with its positive definite
scalar manifold ${\cal M}$. 
If the actions (\ref{5dActionNonSusy}), (\ref{EucActIndef}) and (\ref{MinkAct})
are parts of supersymmetric actions, then $N_{ij}$ is subject to additional
constraints. 

In \cite{EucI}, \cite{Mohaupt:2009iq} it was shown that instanton solutions of
(\ref{EucActIndef}), where the harmonic functions are taken to be 
of the single centered type,
\[
H_i(r) = A_i + \frac{B_i}{r^2}
\]
lift to solitonic solutions with electrical charge $Q_i \propto B_i$
and mass
\[
M = \sigma^i(\infty) Q_i \;.
\]
Here the mass is defined as the integral of the energy density,
given by the component $T_{00}$ of the energy momentum tensor, over space.
In comparison to the amplitude calculation, the role of the boundary 
is played by an asymptotic three-sphere at infinity, and not by 
hyperplanes at infinite Euclidean time. The center $r=0$ of the 
harmonic function also needs to be treated as a boundary. Proper
solitons correspond to solutions where the centers do not contribute,
and it was shown in \cite{Mohaupt:2009iq} that this is the case if the
Hesse potential is a homogeneous function of non-positive degree. 
For Hesse potentials of positive degree the contribution from the
center is infinite, so that such theories do not admit solitonic solutions of
this type.\footnote{As we will see below, this is equivalent to the
statement that the reduced Euclidean theory does not have instantons,
because the saddle point solution has infinite action.}
In the language of $p$-branes, the center $r=0$ is the location 
of a $0$-brane of the five-dimensional theory, which becomes a
$(-1)$ brane upon reduction over time. The tensions of $p$-branes
are related by dimensional reduction, and for $0$-branes and $(-1)$ 
branes the mass and action respectively play the role of the tension.
Thus we expect that the mass of the soliton equals the action of the
instanton from which it was generated.\footnote{There is a numerical 
constant proportional to the volume of the compactified direction which we 
take to be 1 here. Our normalization of the five-dimensional electric charge 
is such that it is equal to the axionic charge.} Since the instanton action is
\[
S_{\rm inst} = b^i (\infty) Q_i \;,
\]
we find that $M=S_{\rm inst}$ requires $b^i(\infty) = \sigma^i(\infty)$, or
$C^i=0$, which is the same condition as we found when describing 
instanton solutions in terms of the dual tensor field theory. Note that
like dualization dimensional lifting has the effect to convert the axionic
shift symmetries into proper gauge symmetries. The soliton mass cannot
depend on $b^i$ since this would make it gauge dependent. 

For completeness let us mention that there are more general
solutions based on multi-centered harmonic functions
\[
H_i({\bf x}) = A_i + \sum_{a=1}^N \frac{B_{i,a}}{|{\bf x} - 
{\bf x}_a|^2} \;,
\]
which describe static configurations of solitons. For such solutions
the total energy is the sum of the masses of the solitions, thus
reflecting the saturation of a Bogomol'nyi bound. The relation 
between this Bogomol'nyi bound and the one for instantons will be explained
later for the special case of supersymmetric (BPS) solutions.

\subsection{Lifting with gravity}

Solutions of the indefinite Euclidean action (\ref{EucActIndef})
which satisfy the extremal instanton ansatz $\partial_m \sigma^i = 
\pm \partial_m \beta^i$ remain solutions, without modification, if
we add a four-dimensional Euclidean Einstein-Hilbert term
\begin{equation}
\label{4dActionGeneral}
S = \frac{1}{2} \int d^4 x \sqrt{g_{(4)}} \left( - R_{(4)} + N_{ij} 
\partial_m \sigma^i \partial^m \sigma^j - N_{ij} 
\partial_m b^i \partial^m b^j \right) \;.
\end{equation}
The resason is that (\ref{ExtInstAnsatz}) implies that the
energy momentum tensor vanishes, so that 
the four-dimensional Euclidean Einstein equations are
solved by a flat metric $g_{mn} = \delta_{mn}$ (or more generally
by a Ricci-flat metric). The resulting solutions of (\ref{4dActionGeneral})
can be lifted to extremal black hole solution of a five-dimensional
action of the following form \cite{Mohaupt:2009iq}:
\begin{equation}
\label{5dActionGeneral}
S = \int \sqrt{g_{(5)}} d^5 x \left( \frac{1}{2} R_{(5)} 
- \frac{3}{4} a_{ij}(h) \partial_\mu h^i \partial^\mu h^j 
- \frac{1}{4} a_{ij}(h) F^i_{\mu \nu} F^{j|\mu \nu} \right)\;.
\end{equation}
The presence of gravity complicates the dimensional lifting/reduction
considerably. In the following we give a concise summary, and refer
to \cite{EucI}, \cite{Mohaupt:2009iq} for more details.
The space-time metrics are related by
\[
ds_{(5)}^2 = - e^{2\tilde{\sigma}} (dt + A_m dx^m)^2 + e^{-\tilde{\sigma}}
ds_{(4)}^2 \;,
\]
where $\tilde{\sigma}$ is the Kaluza Klein scalar and $A_m$ is 
the Kaluza-Klein vector. When lifting instanton solutions,
then $A_m=0$ and $ds_{(4)}^2 = \delta_{mn} dx^m dx^n$, 
so that we obtain the line element of an extremal static black hole solution 
(which can be single or
multi-centered):
\[
ds_{(5)}^2 = - e^{2\tilde{\sigma}} dt^2 + e^{-\tilde{\sigma}}
\delta_{mn} dx^m dx^n\;.
\]
The only non-trivial component of the five-dimensional metric
is the Kaluza-Klein scalar, which is obtained by solving the
four-dimensional scalar equations of motion. The Kaluza-Klein 
vector can be truncated out consistently, because we only consider
non-rotating solutions. Since the metric
contributes one four-dimensional scalar, we need to start with
$n-1$ real scalars and $n$ gauge fields in five dimensions to
obtain a scalar action of the form (\ref{4dActionGeneral})
with $2n$ real scalars. This can be implemented by adapting
a construction well known form five-dimensional vector 
multiplets \cite{Gunaydin:1983bi,Mohaupt:2009iq}.
Instead of working with $n-1$ independent five-dimensional scalars,
one uses $n$ five-dimensional scalars $h^i$ which are subject to
the constraint
\[
\hat{\cal V}(h) = 1 \;.
\]
The constraint 
eliminates one independent degree of freedom. In supergravity
the function ${\cal V}(h)$ is called the prepotential, and must
be a homongeneous cubic polynomial. If we do not require supersymmetry
we can relax this condition and admit any prepotential which is a 
homogeneous function of arbitrary degree $p$. As in supergravity
we require that  
the scalar metric $a_{IJ}(h)$ is Hessian, with Hesse potential
\[
{\cal V}(h) = - \frac{1}{p} \log \hat{\cal V}(h) \;.
\]
Such a Hesse potential is almost, but not quite, homogeneous of
degree zero, and its $k$-th derivatives are homogeneous functions
of degee $-k$. In particular, the scalar metric $a_{ij}$ is homogeneous
of degree $-2$. This feature is crucial in showing that upon 
dimensional reduction (\ref{5dActionGeneral}) becomes
(\ref{4dActionGeneral}), plus terms which are not relevant for
static, purely electric solutions. The key step is to combine
the constrained five-dimensional scalars $h^i$ with the 
Kaluza-Klein scalar $\tilde{\sigma}$ into $n$ unconstrained
real four-dimensional scalars
\begin{equation}
\label{4dScalars5d}
\sigma^i = e^{\tilde{\sigma}} h^i \;.
\end{equation}
One can then use the homogeneity properties of $a_{ij}$ 
to obtain (\ref{4dActionGeneral}). The scalar metric
$N_{ij}(\sigma)$ is proportional to $a_{ij}(h)$.

It has been shown in \cite{Mohaupt:2009iq} that by 
using instanton solutions based on single and multi-centered
harmonic functions one obtains single and multi-centered extremal
black hole solution in five-dimensions, which share all essential
features of BPS black holes of five dimensional vector multiplets.
In particular, one obtains global, algebraic attractor equations, 
which express the solution in terms of harmonic functions, by 
rewriting (\ref{DualScalars}) in terms of five-dimensional variables.
Using the five-dimensional version of the ADM mass formula, one
finds
\[
M_{ADM} = \frac{3}{2} \oint d^3 {\bf x} n^m e^{-\tilde{\sigma}}
\partial_m \tilde{\sigma} \;,
\]
where the integral is over an asymptotic three-sphere.\footnote{As for
solitons based on Hesse potentials which are homogeneous functions of 
non-positive degree, there
are no contributions from the centers.}
If one expresses the instanton action in terms of five-dimensional
quantities, one obtains
\[
S_{\rm inst} = \frac{3}{2} \oint d^3 {\bf x} n^m 
\partial_m \tilde{\sigma} \;.
\]
While the integrands of the surface integrals for $M_{ADM}$ and $S_{\rm Inst}$ 
are different, the factor $e^{-\tilde{\sigma}}$
does not contribute, due to its fall off
behaviour for black hole solutions \cite{Mohaupt:2009iq}. Therefore
ADM mass and instanton action agree, provided that we impose 
$\sigma^i(\infty) = b^i(\infty)$ as discussed before.

\section{Instantons solutions for Euclidean $N=2$ vector multiplets}

We now turn to a particular class of examples, instanton solutions
for four-dimensional Euclidean $N=2$ vector multiplets and the
corresponding solitonic solutions for five-dimensional vector multiplets
and black hole solutions for five-dimensional supergravity with 
vector multiplets. The full actions contain of course further terms
besides those in (\ref{EucActIndef}), (\ref{5dActionNonSusy}),
(\ref{5dActionGeneral}), but for static purely electric backgrounds
these additional terms are not relative. As far as solving the
field equations is concerned, we are just dealing with a particular
subclass of the models considered previously. Therefore we focus
on the additional features due to supersymmetry. We show
how the extremal instanton ansatz can be derived by imposing 
a Euclidean BPS condition, and verify explicitly 
that the resulting solutions
are supersymmetric from both the four-dimensional and the five-dimensional
point of view.

\subsection{The supersymmetry algebra}

We start with the minimal five-dimensional supersymmetry
algebra, from which the four-dimensional Euclidean
supersymmetry algebra can be obtained by dimensional reduction
over time \cite{EucI}. All fermions, including the supercharges
and the supersymmetry transformation parameters are taken to 
be symplectic Majorana spinors.
If $\lambda^i$, where $i=1,2$, is a pair of
complex spinors, then the symplectic Majorana condition is
\[
(\lambda^i)^* = C \gamma_0 \epsilon_{ij} \lambda^j \;.
\]
Here $C$ is the charge conjugation matrix, which satisfies
\begin{equation}
\label{C-matrix}
C^T = - C \;,\;\;\; \gamma^{\mu T} =  C \gamma^\mu C^{-1} \;.
\end{equation}
For definiteness we take $C$ to be real, $C=C^*$.
Here and in the following we usually supress spinor indices
$\alpha, \beta=1,\ldots, 4$:
\[
\lambda^i = (\lambda^i_\alpha) \;,\;\;\;
\gamma^\mu = (\gamma^{\mu \;\beta}_{\alpha}) \;,\;\;\;
C= (C^{\alpha \beta})\;,\;\;\;
C^{-1} = ( C^{-1}_{\alpha \beta}) \;.
\]
The symplectic Majorana condition can be
imposed in five-dimensional Minkowski space, four-dimensional
Minkowski space and four-dimensional Euclidean space, because
in all these cases the charge conjugation matrix can be 
chosen to satisfy (\ref{C-matrix}) \cite{EucI}.

The minimal five-dimensional supersymmetry algebra takes
the form
\begin{equation}
\{ Q_{i \alpha}, Q_{j \beta} \} = - \frac{1}{2} \epsilon_{ij}
(\gamma^\mu C^{-1})_{\alpha \beta} P_\mu \;.
\end{equation}
The indices $i=1,2$ are
raised and lowered with $\epsilon_{ij}$ and its inverse,\footnote{We
use the NW-SE convention, see \cite{EucI} for more details on our
conventions.}  and $\mu, \nu
=0,1,2,3,5$ are five-dimensional Lorentz vector
indices. The five-dimensional supersymmetry algebra has
the R-symmetry group $SU(2)_R$, under which symplectic Majorana
spinors transform as a doublet. 

The four-dimensional Euclidean supersymmetry algebra is
obtained by dimensional reduction over time. If we restrict
ourselves to states on which $P_0$ operates trivially,
we obtain 
\begin{equation}
\{ Q_{i \alpha}, Q_{j \beta} \} = - \frac{1}{2} \epsilon_{ij}
(\gamma^m C^{-1})_{\alpha \beta} P_m \;,
\end{equation}
where $m=1,2,3,5$ are the four Euclidean directions. The
corresponding Clifford algebra is generated by 
$\gamma^m$. Observe that $\gamma^0$ now plays the role
of a `chirality operator', and assumes the role played
by $\gamma_5$ in four-dimensional Minkowski space. The
R-symmetry group is enhanced to $SO(1,1)_R \times SU(2)_R$ \cite{EucI}.
The non-compact abelian factor is generated
by $\gamma_0$ and acts chirally:
\begin{equation}
Q_i \rightarrow e^{-i \gamma^0 \phi} Q_i \;,
\end{equation}
where $\phi$ is real. 
Since $\gamma_0$ is anti-Hermitean, $-i\gamma_0$ has
real eigenvalues $\pm 1$, and the R-symmetry group
$SO(1,1))_R$ acts by chiral scale transformations.

In four-dimensional Minkowski space $\gamma_0$ is replaced
by $\gamma_5$, which is Hermitean. The resulting R-symmetries
act by
\[
Q_i \rightarrow e^{i \gamma_5 \phi} Q_i
\]
Since $i\gamma_5$ has eigenvalues $\pm i$, these are chiral
phase transformations, and the resulting R-symmetry group
is $U(1)_R$.

To discuss BPS states, we need to add central charges. 
The five-dimensional algebra admits are real central charge
$R$:
\begin{equation}
\{ Q_{i \alpha}, Q_{j \beta} \} = - \frac{1}{2} \epsilon_{ij}
(\gamma^\mu C^{-1})_{\alpha \beta} P_\mu  + \frac{i}{2}
\epsilon_{ij} C^{-1}_{\alpha \beta} R \;.
\end{equation}

The four-dimensional Euclidean supersymmetry algebra is again
obtained by reduction over time, and this time we keep $P_0$
which becomes the second real central charge:
\begin{equation}
\{ Q_{i \alpha}, Q_{j \beta} \} = - \frac{1}{2} \epsilon_{ij}
(\gamma^m C^{-1})_{\alpha \beta} P_m  - \frac{1}{2} \epsilon_{ij}
(\gamma^0 C^{-1})_{\alpha \beta} P_0 
+ \frac{i}{2}
\epsilon_{ij} C^{-1}_{\alpha \beta} R \;.
\end{equation}
Similarly, when reducing with respect to the extra
space direction $\mu=5$, the operator $P_5$ becomes
a central charge, which is usually combined with $R$ into
a complex central charge.

\subsection{BPS states}

A systematic way to obtain the BPS states related to a
supersymmetry algebra is to require that the supersymmetry 
transformation parameters form a zero eigenvector of the
Bogomol'nyi matrix\footnote{We refer to \cite{Townsend:1997wg} for
a review of this method.}, which is the matrix 
formed by the anticommutators of supercharges,
\begin{equation}
\{ Q_{i \alpha}, Q_{j \beta} \} \epsilon^{j \beta} =0 \;.
\end{equation}
This eigenvalue problem is the integrability condition 
resulting from imposing that the BPS state is invariant under
the supersymmetry transformation generated by 
$\epsilon^{j\beta}Q_{j\beta}$, and 
the eigenvectors $\epsilon^{j\beta}$ are the transformation parameters
of the supersymmetry transformation which leaves the BPS state
invariant.

Let us first consider the five-dimensional case where 
this relation becomes 
\begin{equation}
\left( 
- \frac{1}{2} \epsilon_{ij}
(\gamma^\mu C^{-1})_{\alpha \beta} P_\mu  + \frac{i}{2}
\epsilon_{ij} C^{-1}_{\alpha \beta} R
\right) \epsilon^{j \beta} = 0 \;
\end{equation}
upon using the algebra. If we impose that the BPS state
is massive, we can go to its rest frame where 
$P_m =0$. Then $P_0=M$ is the mass and we have:
\begin{equation}
\left( 
- \frac{1}{2} \epsilon_{ij}
(\gamma^0 C^{-1})_{\alpha \beta} M  + \frac{i}{2}
\epsilon_{ij} C^{-1}_{\alpha \beta} R
\right) \epsilon^{j \beta} = 0 \;.
\end{equation}
We use $\epsilon_{ij}$ to lower the $SU(2)_R$ index, apply
$C$ to the equation and use that $C\gamma^0 C^{-1} = \gamma^{0T}$:
\begin{equation}
- M \gamma^{0\;\;\alpha}_{\beta} \epsilon_{i}^\beta  + i R 
\epsilon_i^\alpha =0 \;.
\end{equation}
BPS states saturate the Bogomol'nyi bound $M\geq |R|$, 
therefore $M=\pm R$. Since $\gamma_0$ can always be chosen
to be either symmetric or antisymmetric, 
we obtain
\begin{equation}
i \gamma^{0\alpha}_{\;\;\;\; \beta} \epsilon_{i}^\beta  = \pm \epsilon_i^\alpha \;.
\end{equation}
Thus the `Killing spinor' corresponding to a BPS state must
be an eigenstate of $i \gamma^0$. Since the eigenvalues $\pm 1$
of $i\gamma^0$
are two-fold degenerate, we obtain four (real) Killing spinors and
a state satisfying $M=\pm R$ is invariant under one half of the
supersymmetry algebra. 

This condition still makes sense in the Euclidean supersymmetry
algebra, which is obtained by dimensional reduction over time.
Only the interpretation changes: the five-dimensional mass
$P_0=M$ is a central charge from the four-dimensional Euclidean
point of view, and therefore $M=|R|$ is a relation between the
two central charges of the algebra.
In supergravity the central charge is related
to the electric charge of the soliton. Upon dimensional reduction,
this becomes the `instanton charge', which in the models we will
consider is the axionic charge. The central charge $M$ should
also have a physical interpretation, the natural candidate being
the instanton action.
Then the BPS relation $M=|R|$ becomes a 
relation between instanton action and instanton charge 
from the four-dimensional Euclidean point of view. 
Since $\gamma_0$
plays the role of the chirality operator in the Euclidean 
Clifford algebra, the condition imposed on the supersymmetry 
parameters is a chirality condition. As we will see below, this
implies that bosonic BPS field configuration satisfy 
(\ref{ExtInstAnsatz}), which should be viewed as a self-duality
condition.

\subsection{The Euclidean $\mathcal{N}=2$ Vector Multiplet}

Off shell four-dimensional Euclidean vector multiplets were
constructed in \cite{EucI}, to which we refer for details.
In the following we use the same conventions as in \cite{EucI}
and label vector multiplets, and, hence, scalars by captial
indices $I,J, \ldots = 1, \ldots, n$. In previous sections
these indices were denoted $i,j, \ldots$, which in this section 
we reserve for the $SU(2)_R$ indices. 

The field
content of a Euclidean off-shell vector multiplet is
as follows:
\[
(X^I, \lambda^{Ii}, A^I_m | Y^{ij I}) \;.
\]
Here $X^I$ are scalars, $\lambda^{Ii}$ fermions, $A^I_m$
gauge fields, and $Y^{ijI}$ are auxiliary fields. While the
scalars are coordinates on the scalar manifold ${\cal E}$, the other fields
carrying a manifold index are sections of its tangent bundle
$T{\cal E}$.

As we saw above, when going from Minkowski to Euclidean 
signature the $R$-symmetry group of the $N=2$ supersymmetry
changes from $SU(2)_R \times U(1)_R$ to $SU(2)_R \times SO(1,1)_R$.
This was already observed in \cite{Zumino:1977yh} who also observed
that this changes the metric of the scalar manifold for vector multiplets.
In \cite{EucI} this was explored systematically. The generator
of the abelian factor of the $R$-symmetry group acts isometrically 
on the scalar manifold. While
a $U(1)_R$ factor implies that the scalar manifold carries a complex
structure, its non-compact form $SO(1,1)_R$ implies that the complex
structure is replaced by a para-complex structure. In the following
we elaborate on the brief discussion of para-complex geometry
given in Section 3, and refer to \cite{EucI,EucIII} for a comprehensive
account.

Concerning the scalar fields $X^I$, which are interpreted as coordinates
on ${\cal E}$, there are two possible descriptions. One option, which
we already used in Section 3 is to take them to be para-complex
coordinates 
\begin{equation}
\label{PCcoordinates}
X^I = \sigma^I + e b^I \;,
\end{equation}
which stresses the analogy with the complex fields $X^I=\sigma^I + i b^I$
used in the Minkowski version of the theory. This is natural, but 
there is alternative choice, which avoids using the para-complex
unit $e$, namely `null' or `lightcone' coordinates
\begin{equation}
\label{Acoordinates}
X^I_{\pm} = \sigma^I \pm b^I  \;.
\end{equation}
This alternative description is not available in Minkowski signature,
because a complex structure has imaginary eigenvalues $\pm i$. Therefore
its eigenvectors cannot be real, and one is forced to work with the
complexified tangent bundle. 
In constrast, a para-complex structure has real eigenvales $\pm 1$
and real eigenvectors. The null coordinates go along the integral
curves of these eigenvectors and thus provide adapted coordinates.
Therefore one can work with adapted real coordinates and 
the real tangent bundle $T{\cal E}$, thus avoiding to make use of the 
para-complex unit $e$. However, we prefer to use para-complex 
coordinates, and to use the para-complexified tangent bundle 
$T{\cal E}_C$ because of the close analogy with the complex case.
In particular, formulae written in local coordinates can be 
translated systematically from Minkowski to Euclidean space
by substituting $i \rightarrow e$, for all factors of $i$ related
to the complex structure of the scalar target space. A more 
geometric way of expressing this is that we replace the 
complex scalar manifold  ${\cal M}$ and its complexified 
tangent bundle $T{\cal M}_{\mathbbm{C}}$ by the para-complex
manifold ${\cal E}$ and its para-complexified tangent bundle
$T{\cal E}_C$. 

Besides the scalars $X^I$, the vector multiplets contain 
spinors $\lambda^{iI}$, vector fields $A^I_m$ and auxiliary 
fields $Y^I_{ij}$. All these fields carry a manifold index $I$ and
can therefore be interpreted as sections of the tangent bundle 
of ${\cal E}$, either the real tangent bundle $T{\cal E}$ or its
para-complexified version $T{\cal E}_C$. Since we prefer the
para-complex version, we need to comment on some of its 
particular features. 

The spinors $\lambda^{iI}$ are sections of the product bundle
$T{\cal E}_C \times S$, where $S$ is the spin bundle. Since 
spinors are complex, $S$ carries a complex structure $\tilde{I}$.
When going from Minkowski to Euclidean signature the underlying
Clifford algebra changes signature, but $S$ is complex in both 
cases.\footnote{
To avoid confusion,  we emphasize that the complex structure of
$S$ is not replaced by a para-complex structure when going to
Euclidean signature.}
On the product bundle the para-complex structure $J$ of ${\cal E}$
acts as $J \otimes \mathbbm{1}$, while $\tilde{I}$ acts as
$\mathbbm{1} \otimes \tilde{I}$. It is manifest that both 
structures commute:
\[
(\mathbbm{1} \otimes \tilde{I})(J \otimes \mathbbm{1}) =  
{J} \otimes \tilde{I}    =
(J \otimes \mathbbm{1})(\mathbbm{1} \otimes \tilde{I}) \;,
\]
because they operate on different factors of the product
$T{\cal E}_C \times S$. In terms of local coordinates,
this amounts to $ie=ei$. 

When working in local coordinates it is not always 
obvious how to translate formulae of the Minkowski theory
to formulae in the Euclidean theory, because in the 
Minkowski theory there are two types of `$i$', those related 
to the complex structure of the target space, and those related
to the spinor representation. Only the first ones are replaced
by $e$'s, and any ambiguity has to be resolved by identifying
which complex structure is behind a given factor of $i$ in the
Minkowski theory. \footnote{When working with complexified 
axions, there is yet another complex structure, which needs
to be distinguished from the two discussed here. See the appendix 
of \cite{Mohaupt:2009iq} for a formalism which helps to disentangle 
these complex structures when working in local coordinates.}

After this general discussion, we list some formulae that we
will need later on.
Within the para-complex formalism, the  spinors $\lambda^{iI}$
can be decomposed into two para-complex `chiral' parts
\[
\lambda^{iI} = \lambda^{iI}_+ + \lambda^{iI}_- \;,
\]
where
\[
\lambda^{iI}_\pm = \Gamma_{\pm} \lambda^{iI}\;,\;\;\;
\Gamma_{\pm} = \frac{1}{2} ( \mathbbm{1} \pm e (-i) \gamma^0 ) \;.
\]
To understand the geometrical, coordinate independent meaning of 
these formulae, first note that the symplectic Majorana condition 
is a reality condition on $S$, not on $T{\cal E}_C$. Thus it involves
the $i$'s but not the $e$'s. Second, we take the fermions to be
sections of the para-complexified tangent bundle, which can be 
decomposed into a para-holomorphic subbundle with eigenvalue $+e$
and an anti-para-holomorphic subbundle with eigenvalue $-e$, 
with respect to the para-complex structure. The above projection 
acts on both factors of the product $T{\cal E}_C \times S$ 
simultanously and has para-complex eigenvalues $\pm e$. We will see
below that this is convenient, because when using the para-complex spinors
$\lambda^i_{\pm}$ the supersymmetry transformations take exactly the 
same form in  Minkowski and Euclidean signature.

When using adapted coordinates (\ref{Acoordinates}),
there are no $e$'s in the formulae but 
there is an analogous chiral decomposition
\[
\lambda^{iI} = \xi^{iI}_+ + \xi^{iI}_- \;,
\]
where 
\[
\xi^{iI}_{\pm} = \Gamma_{\pm} \xi^{iI} \;,\;\;\;
\Gamma_{\pm} =
\frac{1}{2} ( \mathbbm{1} \pm (-i) \gamma^0 )  \;.
\]
The projector is now real valued, and 
looks similar to 
the normal four-dimensional chiral projection,
with $\gamma_5$ replaced by $\gamma^0$.

Each vector multiplet also contains a gauge
field $A^I_m$, with corresponding field strength 
$F^I_{mn} = 2 \partial_{[m} A^I_{n]}$.
When working in para-complex coordinates, we define the
self-dual and anti-self-dual parts of the fields strengths
as elements of the para-complex tangent space:
\[
F^I_{\pm|mn} = \frac{1}{2} ( F^I_{mn} \pm \frac{1}{e}
\tilde{F}^I_{mn} )\;.
\]
As for the fermions, the reason for introducing 
para-complex fields is that the supersymmetry transformations
take the same form as in Minkowski signature.
When using adapted coordinates, one would instead take the standard
Euclidean decomposition
\[
F^I_{\pm|mn} = \frac{1}{2} ( F^I_{mn} \pm \tilde{F}^I_{mn} ) \;.
\]

Finally, the balance of the off-shell degrees of freedom
is provided by three real scalar fields, which form a 
triplet under $SU(2)_R$. They are organised into a 
symmetric tensor, 
\[
Y^{ij I} = Y^{ji I}
\]
which is subject to the reality condition
\[
(Y^{ijI})^* = Y^I_{ij} = \epsilon_{ik} \epsilon_{jl} Y^{kl I} \;.
\]

In the para-complex formalism, the supersymmetry transformations
take the following form \cite{EucI}:
\begin{eqnarray}
\delta X^I &=& i\bar{\epsilon}^i_+\lambda^I_+ \\ \delta\bar{X}^I&=&i\bar{\epsilon}_-\lambda^I_- \\
\delta\lambda^{iI}_+&=&-\frac{1}{4}\gamma^{mn}F^I_{-mn}\epsilon^i_+ - \frac{i}{2}\slashed{\partial}X^I\epsilon^i_- -Y^{ijI}\epsilon_{+j} \label{eq:SUSY}\\
\delta\lambda^{iI}_-&=&-\frac{1}{4}\gamma^{mn}F^I_{+mn}\epsilon^i_- - \frac{i}{2}\slashed{\partial}\bar{X}^I\epsilon^i_+ -Y^{ijI}\epsilon_{-j}\\
\delta A^I_m&=&\frac{1}{2}\left(\bar{\epsilon}_+\gamma_m\lambda^I_-+\bar{\epsilon}_-\gamma_m\lambda^I_+ \right) \\ \delta Y^{ijI}&=&-\frac{1}{2}\left(\bar{\epsilon}^{\left(i\right.}_+\slashed{\partial}\lambda^{j\left.\right)I}_- +\bar{\epsilon}^{\left(i\right.}_-\slashed{\partial}\lambda^{\left.j\right)I}_+\right) \;.
\end{eqnarray}
The supersymmetry transformations of the Minkowski signature theory 
take precisely the same form, with para-complex quantities replaced 
by complex quantities as explained before.\footnote{When comparing 
to the literature one needs to take into account that most of the $N=2$
literature uses Majorana spinors, which do not exist in Euclidean
signature. Instead we use symplectic Majorana spinors, which exist 
in both signatures. See \cite{EucI} for the details.} 
Supersymmetry acts chirally in the sense that para-holomorphic
scalars transform into positive chirality spinors which transform
into anti-selfdual field strength.

The vector multiplet Lagrangian is gauge invariant, and therefore
it does not involve the gauge field $A^I_m$ directly, but only
the gauge invariant field strength $F^I_{mn}$. The anti-self-dual 
part of the field strength is part of a restricted chiral
multiplet
\[
(X^I, \lambda^{Ii}_+, F^{I}_{- |mn} | Y^{ij I} ) \;,
\]
while the self-dual part 
belongs to the complex (and para-complex) conjugated multiplet.
Since supersymmetry acts chirally it 
does not mix the above
restricted chiral multiplet with its complex conjugate.

\subsection{Supersymmetric field configurations}

Our goal is to identify the BPS field configurations of Euclidean
vector multiplets where the only excited fields are scalars.
Therefore we set the fermions and the gauge fields to zero in 
(\ref{eq:SUSY}), and obtain the following condition on the scalars
$X^I$ and the auxiliary fields $Y^{ijI}$:
\begin{eqnarray}
\delta\lambda^{iI}_+&=&
- \frac{i}{2}\slashed{\partial}X^I\epsilon^i_- -Y^{ijI}\epsilon_{+j} =0 
\\
\delta\lambda^{iI}_-&=& - \frac{i}{2}\slashed{\partial}\bar{X}^I\epsilon^i_+ 
-Y^{ijI}\epsilon_{-j} = 0 \\
\delta Y^{ij I} &=& 0 \;. 
\end{eqnarray}
It is consistent to set $Y^{ijI}=0$, which is in fact
the equation of motion of the auxiliary fields in a bosonic background.
The remaining condition on the scalar is
\begin{equation}
-\frac{i}{2}\slashed{\partial}X^I\epsilon^i_-=0
\end{equation}
together with its complex conjugate.\footnote{Here and in the following
`complex conjugate' as a short hand for taking the para-complex
conjugation of scalars and tangent vectors, and the complex conjugate
of spinors.}
Like the fermions $\lambda^{iI}$, the supersymmetry transformations
parameter $\epsilon^i$ are symplectic Majorana spinors, and 
they have been decomposed into para-complex chiral components
\[
\epsilon^i = \epsilon^i_+ + \epsilon^i_- \;,
\]
\[
\epsilon^i_\pm = \frac{1}{2} ( \epsilon^i 
\pm e (-i) \gamma_0 \epsilon^i )  \;.
\]
From our analysis of the algebra we have seen that the
BPS condition for a point-like object (an instanton in
four dimensions, a soliton in five dimensions) 
is either
\begin{equation}
\label{EpsilonMinus}
\gamma_0 \epsilon^i = - i \epsilon^i \;,
\end{equation}
or 
\begin{equation}
\label{EpsilonPlus}
\gamma_0 \epsilon^i = i \epsilon^i \;.
\end{equation}
Each of these choices reduces the number of independent
supersymmetry parameters from 8 to 4. Therefore invariant
field configurations have 4 Killing spinors and 
are `$\frac{1}{2}$-BPS'. To work out the resulting condition
on the scalars $X^I$, let us make the choice (\ref{EpsilonMinus})
for definiteness. Then 
\begin{equation}
\epsilon^i_\mp = \frac{1}{2} ( 1\mp  e) \epsilon^i \;,
\end{equation}
and the constraint on scalar fields becomes
\begin{equation}
\slashed{\partial} (\sigma^I \pm e b^I ) ( 1 \mp e ) \epsilon^i =0 \;.
\end{equation}
We want to find solutions which do not require to set the
scalars to constant values, which would lead to a fully supersymmetric
ground state solution. While this would be impossible in 
Minkowski signature with its complex target space, in Euclidean
signature with its para-complex geometry we can make use of the
identity
\[
(1+e)(1-e) = 0 \;.
\]
This provides us with a non-trivial solution,
\[
\slashed{\partial} \sigma^I  = \slashed{\partial} b^I \;.
\]
If we choose to impose the other possible constraint
(\ref{EpsilonPlus}) on the supersymmetry parameters, we obtain
$
\slashed{\partial} \sigma^I  = -\slashed{\partial} b^I
$
Combining both cases we see that the condition for a 
scalar field configuration to be 
$\frac{1}{2}$-BPS is:
\[
\slashed{\partial} \sigma^I  = \pm\slashed{\partial} b^I \;.
\]
Since the $\gamma$-matrices are linearly independent, this
is equivalent
to 
\begin{equation}
\partial_m \sigma^I =\pm \partial_m b^I \label{eq:instrelation}
\end{equation}
which is the extremal instanton ansatz (\ref{ExtInstAnsatz}).\footnote{
Remember that the fields $b^I$ were denoted $\beta^i$ in the previous
sections.}

The geometrical interpretation of this condition is that the
solution can only vary along the eigendirections of the para-complex
structure. Looking back at the construction of solutions for
general class of actions we see that we still obtain solutions 
if we relax the condition that the relative sign must be the
same for all values of $I$. However, such solutions, while
extremal, are not BPS. Thus by flipping signs we can generate
non-BPS extremal solutions from BPS solutions. This observation 
agrees with the general pattern observed in the context of 
black holes. The target space metric of our model
\[
ds^2_{\cal E} = N_{IJ}(\sigma) (d \sigma^I d\sigma^J - db^I db^J)
\]
has manifest discrete isometries $R^{I}_{\;J}$ acting by 
\[
b^I \rightarrow R^I_{\;J} b^J = \pm b^I
\]
satisfying
\[
N_{IJ} R^I_{\;k} R^{J}_{\;L} = N_{KL} \;.
\]
Such isometries generate new extremal solutions which are related
to the old ones by rotating the axion charges, as was realized
for black holes in \cite{Ceresole:2007wx,LopesCardoso:2007ky}. 
The BPS condition 
allows to reduce the field equations to gradient flow equations 
driven by the central charge. This feature carries over to extremal
non-BPS solutions with the flow now being driven by a 
`fake superpotential'. For instanton solutions this was discussed
in some detail in \cite{Mohaupt:2009iq}.

\subsection{BPS condition in adapted coordinates \label{BPS-adapted}}

For completeness, let us show explicitly how the same conclusions
can be arrived at when using the adapted real coordinates 
$X^I_{\pm} = \sigma^I \pm b$
instead ot the para-complex coordintes $X^I =\sigma^I + e b^I$. 
Then 
the Euclidean BPS condition for a purely scalar background is
\cite{EucI}:
\begin{eqnarray}
\delta \xi^{iI}_+ &=& - \frac{i}{2} \slashed{\partial} X^I_+ \eta^i_- \;,
\nonumber \\
\delta \xi^{iI}_- &=& - \frac{i}{2} \slashed{\partial} X^I_- \eta^i_+ \;,
\end{eqnarray}
where $\xi^{iI}_\pm = \frac{1}{2} (\mathbbm{1} \pm (-i) \gamma^0) \xi^{iI}$
are the chiral projections (with respect to $\gamma^0$) of the
vector multiplet fermions and 
$\eta^i_\pm = \frac{1}{2} (\mathbbm{1} \pm (-i) \gamma^0) \epsilon^i$
are the chiral projections of the supersymmetry parameters.

Imposing $i\gamma^0 \epsilon^i = \epsilon^i$ implies $\eta^i_+ =0$
and the condition on the scalar fields is
\[
\slashed{\partial} X^I_+ = 0 \Leftrightarrow \partial_m \sigma^I =
- \partial_m b^I \;.
\]
If we impose instead $i\gamma^0 \epsilon^i = - \epsilon^i$, then
$\eta^i_+ =0$ and the condition on the scalars is
\[
\slashed{\partial} X^I_- = 0 \Leftrightarrow \partial_m \sigma^I =
\partial_m b^I \;.
\]
Combining both cases we recover (\ref{eq:instrelation}).

\subsection{Extension to supergravity}

So far our discussion was based on rigid Euclidean vector multiplets.
As discussed in Section 5 we can also couple the sigma model
to gravity and lift instantons to black hole solutions.
We would therefore like to check
that the Euclidean BPS solutions obtained in this section lift
to BPS black holes. One way of doing this would be to work out
the Euclidean supersymmetry transformations for Euclidean 
vector multiplet coupled to supergravity. This has not been done
yet, but at least the bosonic part of the action was worked
out in \cite{EucIII}. We can still use the general fact that 
extremal instanton solutions can be constructed consistently 
taking the Euclidean metric to be flat, because the extremality
condition implies that the energy momentum tensor vanishes.
Therefore we can consistently truncate the bosonic action found
in \cite{EucIII} to its scalar sector. Then the only difference
between the rigid and the local model is the condition 
imposed on the scalar metric $N_{IJ}$. In the rigid case
$N_{IJ}$ must be an affine special para-K\"ahler metric \cite{EucI}.
In terms of local coordinates this means that $N_{IJ}$ has
a para-K\"ahler potential $K(X,\bar{X})$
\[
N_{IJ} = \frac{\partial^2 K(X,\bar{X})}{\partial X^I \partial
\bar{X}^J}
\]
which can be obtained from para-holomorphic prepotential 
$F(X)$ by
\[
K(X,\bar{X}) = -e(X^I \bar{F}_I - F_I \bar{X}^I) \;.
\]
For models obtained by dimensional reduction of five-dimensional
vector multiplets over time, the prepotential is homogeneous
cubic polynomial $F(X) = C_{IJK} X^I X^J X^K$.

In the local case the para-K\"ahler form instead takes the
form 
\[
K(X,\bar{X}) = - \log [-e(X^M \bar{F}_M - F_M \bar{X}^M)] 
\]
where $M,N=0, 1, \ldots, n$. The prepotential depends 
on one additional variable $X^0$, but is required to be
homogeneous of degree two, so that the scalar metric only
depends on $n$ of the $n+1$ variables. For models obtained
by dimensional reduction of five-dimensional vector multiplets,
the prepotential takes the form $F(X) = (X^0)^{-1} C_{IJK} X^I X^J X^K$.

Since the off-shell supersymmetry transformation rules given for 
rigid Euclidean vector multiplets do not depend on the prepotential,
it is clear that we will have the same transformations in 
supergravity if we impose a flat metric. Therefore the condition 
for scalar $\frac{1}{2}$-BPS is not modified. In fact, this type
of reasoning can be applied generally to asymptotically flat 
solutions of supergravity to find the algebraic structure of 
Killing spinors by looking for eigenvectors of the matrix
of supersymmetry anticommutators, see for example \cite{Townsend:1997wg}. 
We can check 
this by comparing our BPS condition to the one 
found for five-dimensional static BPS black holes in \cite{Chamseddine:1998yv},
\[
\epsilon_i = e^{\tilde{\sigma}} \epsilon_{(0)i} \;,
\]
where $\epsilon_{(0)i }$ are constant spinors subject to the
BPS condition $\gamma^0 \epsilon_{(0)i}  = \mp i \epsilon_{(0)i}$.
This is of course the condition that we found earlier by 
looking for an eigenvector of the Bogomol'nyi matrix 
corresponding to massive pointlike BPS state. 
Moreover, for non-rotating, 
purely electric black holes, the condition imposed on the
bosonic fields in five dimensions is \cite{Chamseddine:1998yv}
\[
\partial_m A_0^I =   \pm \partial_m ( e^{\tilde{\sigma} }h^I) \;.
\]
In terms of four-dimensional variables this becomes
\[
\partial_m b^I = \pm \partial_m \sigma^I \;,
\]
which is indeed the same BPS condition as we found for rigid Euclidean
vector multiplets.

Let us finally make a remark on the mass of the solitons and
black holes obtained by lifting BPS instantons. For the general
class of models the relation between soliton mass and instanton
action was discussed previously. Moreover, it was shown in \cite{Mohaupt:2009iq}
that the instanton action and soliton mass are only finite if
there is no contribution from the centers of the harmonic function.
For metrics $N_{IJ}= \partial_I \partial_J H(\sigma)$ with a homogenous
Hesse potential, this requires that Hesse potential is either
homogeneous of negative degree, or is the logarithm of a 
homogenous function of arbitrary degree. For rigid vector multiplets
the Hesse potential and prepotential are related by \cite{EucIII,Mohaupt:2009iq}
\[
F(X) = F(\sigma + e b) = 4 H(\sigma + e b) \;.
\]
Since the prepotential, and, hence, the Hesse potential are homogeneous
of degree three, we conclude that BPS solutions have infinite mass.
Therefore these solutions are not instantons and solitons.

However for local vector multiplets, the four-dimensional prepotential 
$F(X) = (X^0)^{-1} C_{IJK} X^I X^J X^K$ is related to the five-dimensional
Hesse potential \cite{EucIII,Mohaupt:2009iq}
\[
H(\sigma) = - \log C_{IJK} \sigma^I \sigma^J \sigma^K
\]
which is the logarithm of a homogeneous function. Thus in supergravity
Euclidean BPS solutions have finite action and lift to BPS black holes
of finite mass. Finally let us remark that for the Euclidean STU
model, solutions can further be lifted to ten-dimensional 
five-branes \cite{EucIII}.
Therefore these solutions are relevant for understanding five-brane
instantons in heterotic string theory.

\section{Discussion and Outlook}

In this paper we have discussed the relation between different
Euclidean formulations of axions and their implications on 
instantons and solitons. We observed that the instanton action
depends on the boundary values of the axionic fields, reflecting
the breaking of continuous shift symmetries. This feature neither
occurs in the dual tensor version of the theory nor in the 
dimensionally lifted theory. While there is no apparent contradiction
in accepting 
that quantum theories can be `equivalent up to 
zero modes', it becomes an issue as soon as one wants to compute
physical effects due to instantons or solitons in a particular
theory. Then one has to decide whether the physical theory one
wants to study is a theory of axions or of antisymmetric tensor
fields. This is relevant for the physics of wormholes, and
also for the instanton calculus of string theory. 
According to \cite{BurgessKshirsagar:1989} the formation
of wormholes and baby universes only occurs naturally in axionic
theories. Similarly, since the breaking of axionic shift symmetries
is believed to be a non-perturbative effect relevant to 
low energy physics and hence phenomenology, the axionic formulation
should also be fundamental for string effective theories.
However, in practice the dual tensor field formulation
is often used, and the programme of finding all non-perturbative
corrections to hypermultiplets by using string dualities (reviewed in 
\cite{Saueressig:2007gi}) has so
far relied on this. Further progress will require a better 
understanding of instantons within the axionic framework.
We also remark that 
the vector multiplet instantons discussed in this paper correspond
to five-brane instantons and world-sheet instantons of heterotic
string compactifications \cite{EucIII}. It would be interesting to use this
for a direct check of heterotic-type II string duality.

One restriction that we imposed is that all shift symmetries
commute. While this has the advantage that some conceptual 
aspects could be discussed very clearly, it is not the most
general case of interest. The geometry of hypermultiplets provides
an example of a more complicated class of geometries, with 
isometries forming a centrally extended Heisenberg group
instead of an Abelian group. A generalized version of the c-map
\cite{Ferrara:1989ik},
in analogy to the generalized r-map introduced in \cite{AlekCor,Mohaupt:2009iq}
should be the appropriate framework for investigating this larger
class of models. One should then be able to tackle the issue 
raised in \cite{Chiodaroli:2009cz} systematically, namely 
which saddle points are really relevant for the semi-classical
approximation. In the context of generating 
solitonic solutions from instantons, geometries of the type
found in hypermultiplets allow to include magnetic in addition
to electric charge.

While we have investigated the Euclidean formulation of
axions, there are issues with the Euclidean formulation of fermions
as well, in particular in the context of supersymmetric 
theories \cite{Nicolai:1978,vanNieuwenhuizen:1996tv,vanNieuwenhuizen:1996ip}. 
We found that it is natural to complexify the axions, and in supersymmetric
it then suggests itself to complexify all fields. For maximal eleven-
and ten-dimensional supergravities the use of complexified fields
leads to a simplified description of dualities, instantons and 
solitons \cite{Bergshoeff:2007cg}. In the appendix of \cite{Mohaupt:2009iq}
we gave an outline of how to complexify the scalar target spaces 
of the sigma models considered in this paper. For $N=2$ theories
it should be useful to extend this to the whole theory, resulting in 
a `complexified version' of special geometry. 
If one is interested in time-like T-duality transformations
and type-II$^*$ string theories \cite{Hull:1998vg}, 
the use of complexified fields should
also be useful in Minkowski signature.


\end{document}